\documentstyle[preprint,prc,aps,eqsecnum,epsf]{revtex}
\usepackage[dvips]{epsfig}
\tighten
\begin{document}
\draft

\title{Electromagnetic Structure of the Trinucleons}
\author{L.E.\ Marcucci}
\address{Department of Physics, Old Dominion University,
         Norfolk, Virginia 23529}
\author{D.O.\ Riska}
\address{Department of Physics, University of Helsinki, 
        SF-00014 Helsinki, Finland}
\author{R.\ Schiavilla}
\address{Jefferson Lab, Newport News, Virginia 23606 \\
         and \\
         Department of Physics, Old Dominion University,
         Norfolk, Virginia 23529}
\date{\today}

\maketitle

\begin{abstract}
The electromagnetic form factors of the trinucleons
$^3$H and $^3$He are calculated with wave functions
obtained with the Argonne $v_{18}$ two-nucleon
and Urbana IX three-nucleon interactions.
Full account is taken of the two-body currents 
required by current conservation with the $v_{18}$
interaction as well as those associated with 
$N\Delta$ transition currents and the currents of
$\Delta$ resonance components in the wave functions. 
Explicit three-nucleon 
current operators associated with the 
two-pion exchange three-nucleon interaction 
arising from irreducible
$S$-wave pion-nucleon scattering is
constructed and shown to have very little effect on
the calculated magnetic form factors. The calculated
magnetic form factor of $^3$H, and 
charge form factors of both $^3$H and $^3$He
are in satisfactory agreement with the experimental data. 
However, the position of the zero in the magnetic form
factor of $^3$He is slightly underpredicted.
\end{abstract}
\pacs{21.45.+v,25.30.Bf,27.10.+h}

\section{Introduction}
\label{sec:intro}

The electromagnetic form factors of the few-body nuclei, along with the 
deuteron structure functions and cross section for threshold 
electrodisintegration at backward angles, are the observables of choice 
for testing the quality of models for the nuclear interaction and 
the associated current operator, including its exchange
current components. Such testing has become possible by the
development of practical computational methods for 
numerical calculation of 
the wave functions of the few nucleon systems, which correspond
to realistic phenomenological interaction models~\cite{CAR98}.
Employing such wave functions along with the
the two-nucleon
exchange current operators, which are required by the
continuity equation and/or consistency with the interaction
models, e.g., by Poincar\'e invariance, it has become possible
to predict the experimental electron scattering observables of the
few nucleon systems up to momentum transfers of about 2 GeV/c
in an at least qualitatively satisfactory way.

Among the remaining open issues are the need for
quantitative understanding of the form factors of the
trinucleons in the region around and above their first zeros.
While the long standing, unsettled issue of the behavior of the
tensor polarization of the deuteron for momentum
transfers above 3 fm$^{-1}$ appears close to 
settlement by high quality experimental work at the
Thomas
Jefferson National Accelerator Facility, there 
remains a need for potential
model development in the case of the trinucleon systems.
This is partly because of the remaining problem in
quantitative understanding of the trinucleon form factors
at high momentum transfer and partly because of the
resilient open issues concerning the form of the
three-nucleon interaction, which appears to be required
for the understanding of the binding energies of the
light-nuclei ($A\leq$7) ground states~\cite{PUD97}.

Here this question is investigated in several different
ways. First, a numerically extensive calculation of the
electromagnetic form factors of the trinucleons is
presented with high precision variational wave functions,
which correspond to the Argonne $v_{18}$ two-nucleon~\cite{WIR95}
and the Urbana IX three-nucleon~\cite{PUD95} interactions.
In this calculation the two-nucleon exchange current operators
are constructed by the same method as used in the earlier
calculations in Refs.~\cite{SCH89,SCH90}, that employed the
Argonne $v_{14}$ interaction~\cite{WIR84}. Second, the
irreducible three-nucleon exchange current operator, which
corresponds to the best understood part of the two-pion exchange
three-nucleon interaction associated with $S$-wave pion
nucleon scattering on the intermediate nucleon is
constructed, and its matrix elements for the trinucleon
bound states are shown to be very small for momentum
transfer values below 1 GeV/c. Third, a systematic treatment
of $\Delta$-isobar configurations in the trinucleon
ground states is made, and their effect on the trinucleon
form factors are calculated with inclusion of all the
associated and required exchange current operators. 

The calculated magnetic form factor of $^3$H is in fairly
good agreement with the present experimental data once the
exchange current contributions are included. That of $^3$He
agrees less well with the corresponding data at high
values of momentum transfer, which is a consequence of the
underpredicted position of the first zero in the form factor
(it falls at 3.75 fm$^{-1}$, which is below the experimental
range 4.2--4.4 fm$^{-1}$). This result is independent of the
presence or absence of the $\Delta$-isobar configurations
in the wave function model. The problem is likely to have its
origin in a somewhat too weak overall strength of the model for
the isovector part of the exchange current operator
at large momentum transfer.
As this is constructed so as to be consistent with the $v_{18}$
two-nucleon interaction, the ultimate origin of the problem with the
high momentum behavior of the calculated isovector magnetic 
form factor of the trinucleons
may reside with the potential model if not
with some purely transverse exchange current mechanism
that has not been considered. The effect 
of the irreducible
three-nucleon exchange current operator on this
form factor is very small. The finding that the
$\Delta$-isobar configurations have a very small effect on
the trinucleon form factors conforms to earlier
results obtained with other potential models~\cite{PEN90}.

The calculated charge form 
factor of $^3$He agrees very well with the
experimental values over the measured range 
of momentum transfer, with an exception for its highest end. 
In the case of $^3$H quantitative
agreement with the experimental form factor is achieved only up
to the position of the secondary maximum, above which region
the calculated values are too large by factors 2--3. The exchange
current contributions are essential for agreement with the
experimental charge form factors.

This paper is divided into four main sections. Section II
%~\ref{sec:nucl} 
contains a description of the calculation of the electromagnetic form
factors of the trinucleons using a purely nucleonic variational
wave function constructed for the $v_{18}$ model
augmented by the Urbana IX three-nucleon interaction. The
details of the hyperspherical variational model wave function
are given in Subsection II-A. 
%~\ref{subsec:ham+wf} 
Subsection II-B
%~\ref{subsec:1-2ccop} 
contains 
the description of the model for the
electromagnetic current operator including the exchange
current operators. In Subsection II-C
%~\ref{subsec:WTcurr} 
the irreducible three-nucleon
exchange current operator, which corresponds to the 
main nonresonant two-pion
exchange three-nucleon interaction, is 
derived. Finally Subsection II-D 
%~\ref{subsec:eff3} 
contains the form factor results obtained with this
restricted model. In Section III
%~\ref{sec:ND} 
the description of the extended
model wave function and current operators that include
the $\Delta$-isobar configurations in the wave function
is given. The model for the $N\Delta$ transition potential
is described in Subsection III-A
%~\ref{subsec:deltawf} 
and the corresponding
current operators are described in Subsection III-B.
%~\ref{subsec:ND+Dcurr}. 
The calculation
of the form factors in the extended model is outlined in
Subsections III-C and III-D.
%~\ref{subsec:calmff} and~\ref{subsec:mmom+mff}. 
Finally Section IV 
%~\ref{sec:concl} 
contains a concluding discussion.

\section{Trinucleon form factors with nucleonic wave functions}
\label{sec:nucl}
In this section the calculation of the 
elastic form factors 
of the $A$=3 nuclei with wave functions for a 
realistic Hamiltonian model formed of the 
Argonne $v_{18}$ (AV18) two-nucleon~\cite{WIR95} and Urbana IX (UIX) 
three-nucleon~\cite{PUD95} interactions
is described. The calculation employs charge and current operators 
that besides the standard single nucleon components 
also contain two-nucleon components, the 
leading terms of which are constructed 
consistently with the AV18 model. The three-nucleon
exchange current operator, which corresponds to the
two-pion exchange three-nucleon interaction associated
with isospin odd $S$-wave pion rescattering is
also derived and is shown to have only a minor effect
on the trinucleon form factors.

\subsection{The AV18/UIX Hamiltonian and trinucleon wave functions}
\label{subsec:ham+wf}
The AV18 model~\cite{WIR95} is a recent high-quality nucleon-nucleon 
interaction containing explicit charge-symmetry-breaking (CSB) and 
charge-independence-breaking (CIB) terms, as well as a complete treatment 
of the electromagnetic interaction up to order $\alpha^{2}$, $\alpha$ being 
the fine structure constant. It is constructed to fit
the Nijmegen $pp$ and $np$ scattering 
database, low-energy $nn$ scattering parameters, and the 
deuteron binding energy 
with a $\chi^{2}$ per datum close to 1.

The UIX three-nucleon interaction~\cite{PUD95} consists of a 
long-range term due to excitation of an intermediate 
$\Delta$-isobar via two-pion exchange and 
a short-range repulsive phenomenological term, which
simulates the dispersive 
effects which arise upon integrating out $\Delta$-degrees of freedom. 
The strength of this repulsive term is determined by fitting the triton 
binding energy in ``exact'' Green's Function Monte Carlo (GFMC) 
calculations~\cite{PUD97} 
and the equilibrium density of nuclear matter in variational calculations 
based on operator-chain summation techniques~\cite{WIR88}.

Recent GFMC calculations based on the AV18/UIX Hamiltonian model have been 
shown to provide a good description of the low-energy spectra and charge 
radii of nuclei with $A\leq$7~\cite{PUD97}. In particular, 
the calculated binding 
energies of $^{3}$H and $^{3}$He are within a few keV of the experimental 
values (see Table~\ref{tb:binde}). 

In the present work we use trinucleon wave functions obtained by Kievsky 
$et$ $al$.~\cite{KIE93,KIE94,KIE95} with the pair-correlated hyperspherical 
harmonics (PHH) method. Although variational this
method has been refined in the last 
few years so much that it yields results with an accuracy 
comparable to that achieved in recent Faddeev and GFMC calculations, as may
be seen in Table~\ref{tb:binde}. The PHH method is briefly reviewed here, 
for completeness; however, a more thorough discussion of it as well as its 
extensions to describe both the $A$=3 low energy continuum and $A$=4 ground 
state can be found in Refs.~\cite{KIE94,KIE95,VIV95}.

The wave function $\Psi$ of a three-nucleon system with total angular 
momentum $J$$J_{z}$ and total isospin $T$$T_{z}$ can be decomposed as
\begin{equation}
\Psi=\sum_{i=1}^{3}\psi({\bf x}_i, {\bf y}_i) \ ,
\label{eq:fadampl}
\end{equation}
where the amplitude $\psi({\bf x}_i, {\bf y}_i)$ is a function of the 
Jacobi coordinates ${\bf x}_i={\bf r}_j-{\bf r}_k$ and ${\bf y}_i= 
\left({\bf r}_j+ {\bf r}_k-2\,{\bf r}_i\right)/{\sqrt{3}}$, 
$i$,$j$,$k$ being a cyclic permutation of 1,2,3. To ensure the overall 
antisymmetry of $\Psi$, the amplitude $\psi({\bf x}_i, {\bf y}_i)$ is 
antisymmetric with respect to exchange of nucleons $j$ and $k$. In the PHH 
method, it is expressed as~\cite{KIE94,KIE95}
\begin{equation}
        \psi({\bf x}_i, {\bf y}_i)=\sum_{\alpha}
                f_{\alpha}(x_{i}){\Phi}_{\alpha}(x_{i},y_{i}){\cal Y}
                 _\alpha(j,k;i) \ ,
\label{eq:psi1}
\end{equation}
\begin{eqnarray}
{\cal Y}_\alpha(j,k;i)&=&\biggl \{ \Bigl[Y_{\ell_\alpha}(\widehat
           {\bf x}_i)\otimes Y_{L_\alpha}(\widehat {\bf y}_i) 
           \Bigr]_{\Lambda_\alpha}
           \otimes \Bigl[S_\alpha^{jk}\otimes s^i\Bigr]_{S_\alpha} 
           \biggr \}_{JJ_z}
           \Bigl[T_\alpha^{jk}\otimes t^i\Bigr]_{TT_z}\ ,
       \label{eq:caly}
\end{eqnarray}
where each channel $\alpha$ is specified by the orbital angular momenta 
$\ell_\alpha$, $L_\alpha$ and $\Lambda_{\alpha}$, the spin (isospin) 
$S_\alpha^{jk}$ ($T_\alpha^{jk}$) of pair $j$$k$ and the total spin 
$S_{\alpha}$. Orbital and spin angular momenta are coupled, in the LS-scheme, 
to give total angular momenta $J$$J_z$. The  (channel-dependent) correlation 
functions $f_{\alpha}(x_{i})$ are obtained from solutions of two-body 
Schr\"{o}dinger-like equations in channel $j_{\beta}\,l_{\beta}\,
S_{\beta}^{jk}\,T_{\beta}^{jk}$~\cite{KIE93,KIE94}, and take into account the 
strong state-dependent correlations induced by the nucleon-nucleon 
interaction. They improve the behavior of the wave function at small 
interparticle distances. Were it not for their presence, the decomposition 
in Eq.~(\ref{eq:psi1}) would be identical to that in the Faddeev 
scheme~\cite{GLO83}. 

Next, the hyperspherical coordinates $\rho$ and $\phi_{i}$, defined as
\begin{equation}
     \rho= \sqrt{x_i^2+ y_i^2} \ ,\qquad 
      \cos \phi_i= x_i/\rho \ 
      \label{eq:rophi}
\end{equation}
are introduced, and the dependence of ${\Phi}_{\alpha}(x_{i},y_{i})$ on 
$\rho$ and $\phi_{i}$ is made explicit by writing
\begin{equation}
{\Phi}_{\alpha}(x_{i},y_{i})=
                    \sum_{n=0}^{M_\alpha}
                    u_n^\alpha(\rho)
                    Z_{n}^{\alpha}(\phi_i) \ , 
\label{eq:hyperpart}
\end{equation}
\begin{equation}
Z_{n}^{\alpha}(\phi_i)\,=\,N_{n}^{\ell_\alpha,L_\alpha} (\cos{\phi_i})^
{\ell_\alpha}(\sin{\phi_i})^{L_\alpha}\,P_{n}^{\ell_\alpha+\frac{1}{2},
L_\alpha+\frac{1}{2}}(\cos{2\phi_i}) \ ,
\label{eq:jacpol}
\end{equation}
where $N_{n}^{\ell_\alpha,L_\alpha}$ are normalization factors, 
$P_{n}^{\alpha,\beta}$ are Jacobi polinomials and $n$ is a non-negative 
integer, $n=0,\cdots,M_{\alpha}$, where $M_{\alpha}$ is 
the selected number 
of basis functions in channel $\alpha$. The complete wave function is then 
written as
\begin{eqnarray}
        \Psi&=&{\sum_{ijk\,{\rm cyclic}}}\sum_{\alpha}
                    f_{\alpha}(x_{i})  
                    {\cal Y}_\alpha(j,k;i) 
                    \sum_{n=0}^{M_\alpha}
                    u_n^\alpha(\rho)
                    Z^{\alpha}_{n}(\phi_i) \ .
\label{eq:psi}
\end{eqnarray}
The Rayleigh-Ritz variational principle,
\begin{equation}
     \langle\delta_u\Psi | H-E |  \Psi \rangle =0 \ ,
      \label{eq:rr}
\end{equation}
is used to determine the 
hyper-radial functions $u_n^\alpha(\rho)$ in Eq.~(\ref{eq:psi}). Carrying 
out the variation $\delta_u\Psi$ with respect to 
the functions $u_n^\alpha(\rho)$, the following equation is easily derived:
\begin{equation}
{\sum_{ijk\,{\rm cyclic}}}\langle f_{\alpha}(x_{i}){\cal Y}_\alpha(j,k;i)
Z^{\alpha}_{n}(\phi_i) | H-E | \Psi \rangle\mid_{\Omega} =0 \ ,
\label{eq:rr2}
\end{equation}
where $\Omega$ denotes the angular variables $\phi_{i}$, $\hat{\bf x}_{i}$ 
and $\hat{\bf y}_{i}$. Performing the integration over $\Omega$ and 
spin-isospin sums (as implicitly understood by the notation $\langle \cdots 
\rangle\mid_{\Omega}$) leads to a set of coupled second order differential 
equations for the $u_n^\alpha(\rho)$, which is then solved by standard 
numerical techniques~\cite{KIE93,KIE94}.

The binding energy of the $A$=3 nuclei obtained with the PHH method from the 
AV18/UIX Hamiltonian are listed in 
Table~\ref{tb:binde}~\cite{KIE95}. Also listed in 
Table~\ref{tb:binde} are results calculated with converged 
$r$-space~\cite{CHE86} and $p$-space~\cite{GLO95} Faddeev wave functions 
for an older model of the two-nucleon interaction, the Argonne $v_{14}$ 
(AV14)~\cite{WIR84}. The binding energies obtained with the various methods 
are in excellent agreement with each other, typically within 10 keV or less.

\subsection{Nuclear charge and current operators}
\label{subsec:1-2ccop}
A fairly complete decription of the model for the nuclear electromagnetic 
current has been most recently given in Ref.~\cite{CAR98}. 
Here we only review its general structure. The 
nuclear charge and current operators are expanded into a sum of one- and 
two-body terms:
\begin{eqnarray}
\rho({\bf q})&=&\sum_{i}\rho^{(1)}_{i}({\bf q})\,+\,
\sum_{i<j}\rho^{(2)}_{ij}({\bf q}) \ , \label{eq:ch1b} \\
{\bf j}({\bf q})&=&\sum_{i}{\bf j}^{(1)}_{i}({\bf q})+
\sum_{i<j}{\bf j}^{(2)}_{ij}({\bf q}) \ ,
\label{eq:cur1b}
\end{eqnarray}
where ${\bf q}$ is the momentum transfer.
The one-body operators $\rho^{(1)}$ and ${\bf j}^{(1)}$ are given by 
\begin{eqnarray}
\rho_{i}^{(1)}({\bf q})&=&\frac{1}{\sqrt{1+q_{\mu}^{2}/4m^{2}}}
\frac{1}{2}\left[G_{E}^{S}(q_{\mu}^{2})+G_{E}^{V}(q_{\mu}^{2})\tau_{z,i}
\right]e^{{\rm i}{\bf q}\cdot{\bf r}_{i}} \nonumber \\
 & &-\frac{{\rm i}}{8m^{2}}\Bigg[2\,G_{M}^{S}(q_{\mu}^{2})-G_{E}^{S}
(q_{\mu}^{2})+\left[2\,G_{M}^{V}(q_{\mu}^{2})-G_{E}^{V}(q_{\mu}^{2})\right]
\tau_{z,i}\Bigg]{\bf q}\cdot(\bbox{\sigma}_{i}\times{\bf p}_{i})e^{{\rm i}
{\bf q}\cdot{\bf r}_{i}} \ , 
\label{eq:rho1b} \\
{\bf j}_{i}({\bf q})&=&\frac{1}{4 m}\left[G_{E}^{S}(q_{\mu}^{2})+G_{E}^{V}
(q_{\mu}^{2})\tau_{z,i}\right]\,\left\{{\bf p}_{i},e^{{\rm i}{\bf q}\cdot
{\bf r}_{i}}\right\} \nonumber \\
 & &-\frac{\rm i}{4 m}\left[G_{M}^{S}(q_{\mu}^{2})+G_{M}^{V}(q_{\mu}^{2})
\tau_{z,i}\right]\,{\bf q}\times\bbox{\sigma}_{i} e^{{\rm i}{\bf q}\cdot
{\bf r}_{i}} \ , \label{eq:j1bNN} 
\end{eqnarray}
up to terms proportional to $1/m^{2}$, $m$ being the nucleon mass.
Equation~(\ref{eq:rho1b}) includes the leading relativistic 
corrections to single-nucleon charge operator, namely the Darwin-Foldy and 
spin orbit terms.
Here the $G_{E/M}^{S/V}(q_{\mu}^{2})$ are the electric/magnetic ($E/M$) 
isoscalar/isovector ($S/V$) form factors of the nucleon, taken as function 
of the four-momentum transfer
\begin{equation}
q_{\mu}^{2}={\bf q}^{2}-\omega^{2} > 0 \ ,
\label{eq:qm2}
\end{equation}
where the energy transfer $\omega\,=\,\sqrt{{\bf q}^{2}+m_{T}^{2}}\,-\,m_{T}$ 
for elastic scattering on a target of mass $m_{T}$ initially at rest in the 
lab. These form factors are normalized as
\begin{eqnarray}
G_{E}^{S}(0)&=&G_{E}^{V}(0)\,=\,1 \ , \nonumber \\
G_{M}^{S}(0)&=&0.880\, \mu_{N} \ , \nonumber \\
G_{M}^{V}(0)&=&4.706\, \mu_{N} \ , \label{eq:normff}
\end{eqnarray}
$\mu_{N}$ being the nuclear magneton, and their $q_{\mu}$-dependence is 
constrained by analyzing electron-proton and 
electron-deuteron scattering data. 
While the proton electric and magnetic form factors are experimentally 
fairly well known over a wide range of momentum transfers, there is 
significant uncertainty in the neutron form factors, particularly the 
electric one, which are obtained from model-dependent analyses of 
$ed$ data. 
Until this uncertainty in the detailed behaviour of the electromagnetic form 
factors of the nucleon is narrowed, quantitative predictions of 
electro-nuclear observables at high momentum transfers will remain rather 
tentative. We will re-examine this issue in Sec. II-D-1
%~\ref{subsubsec:mff} 
below. 

\subsubsection{The two-body current operator}
\label{subsubsec:2bcurrop}
The two-body current operator can be separated into a model-independent (MI) 
term determined from the interaction (in the present case, the 
charge-independent part of the AV18 model) following a prescription 
originally proposed in Ref.~\cite{RIS89}, 
and a model-dependent (MD) one, associated 
with the $\rho\pi\gamma$ and $\omega\pi\gamma$ electromagnetic couplings. 
Explicit expressions for all these currents have been most recently 
given in Ref.~\cite{VIV96}.

The $\rho\pi\gamma$ and $\omega\pi\gamma$ MD currents are purely transverse 
and therefore unconstrained by the nucleon-nucleon 
interaction. The values of the 
transition form factors $G_{\rho\pi\gamma}(q_{\mu}^{2})$ and 
$G_{\omega\pi\gamma}(q_{\mu}^{2})$ at the photon point are known to be 
$G_{\rho\pi\gamma}(0)\,=\,g_{\rho\pi\gamma}\,=\,0.56$, Ref.~\cite{BER80}, and 
$G_{\omega\pi\gamma}(0)\,=\,g_{\omega\pi\gamma}\,=\,0.68$, 
Ref.~\cite{CHE71-73}, 
from the measured widths of the $\rho\rightarrow\pi\gamma$ and 
$\omega\rightarrow\pi\gamma$ decays, while their $q_{\mu}$-dependence is 
modeled using vector-meson dominance. Monopole form factors at the 
pion and vector-meson strong interaction vertices are introduced to take 
into account the composite nature of nucleons and mesons. The cutoff 
parameters $\Lambda_{\pi}$, $\Lambda_{\rho}$ and $\Lambda_{\omega}$ in these 
form factors are not known. Here we use the values 
$\Lambda_{\pi}\,=\,0.75$ GeV and 
$\Lambda_{\rho}\,=\,\Lambda_{\omega}\,=\,1.25$ GeV obtained from studies 
of the B-structure function of the deuteron~\cite{CAR91}.

The leading MI two-body currents, denoted as pseudoscalar (PS) or 
$\pi$-like and vector (V) or $\rho$-like, are the isovector ones 
associated with the isospin-dependent central, spin-spin, and tensor 
components of the interaction. Their derivation has been given in a number 
of references~\cite{CAR98,RIS85}, and will not be repeated here. 
We only note that: 
i) the PS and V two-body currents have no free 
parameters and, by construction, satisfy the continuity equation with the 
given realistic interaction (here the charge-independent part of AV18 model); 
ii) the continuity equation requires the same form factor be used to 
describe the electromagnetic structure of the hadrons in the longitudinal 
part of the current operator and in the charge operator, while it places 
no restrictions on the electromagnetic form factors which may be used in 
the transverse parts of the current. Ignoring this ambiguity, the form 
factor $G_{E}^{V}(q_{\mu}^{2})$ is used in the PS and V currents operators, 
in line with the ``minimal'' requirements of current conservation.

There are additional two-body currents associated with the momentum 
dependence of the interaction, but their construction is less 
straightforward. A procedure similar to that used to derive the PS and V 
currents has been generalized to the case of the currents from spin-orbit 
components of the interaction~\cite{CAR90}. It consists, in essence, 
of attributing these to exchanges of $\sigma$-like and $\omega$-like mesons 
for the isospin-independent terms, and to $\rho$-like mesons for the 
isospin-dependent ones. The explicit form of the resulting currents, 
denoted as SO, can be found in Refs.~\cite{VIV96,CAR90}. 
The two-body currents from 
the quadratic momentum dependence of the interaction are obtained by minimal 
substitution ${\bf p}_{i}\rightarrow{\bf p}_{i}\,-\,\frac{1}{2}
\left[G_{E}^{S}(q_{\mu}^{2})\,+\,G_{E}^{V}(q_{\mu}^{2})\tau_{z,i}\right]\,
{\bf A}({\bf r}_{i})$, ${\bf A}({\bf r}_{i})$ being the vector potential, 
into the corresponding components. In the case of the AV18 model, the 
$p^{2}$-dependence is via ${\bf L}^{2}$ and $({\bf L}\cdot\bbox{\sigma}_{1}\,
{\bf L}\cdot\bbox{\sigma}_{2}\,+\, {\rm h.c.})$ terms, and the associated 
currents are denoted respectively as LL and SO2~\cite{VIV96,SCH89}.

We note that the SO, LL and SO2 currents are fairly short-ranged, and have
 both isoscalar and isovector terms. Their contribution to isovector 
observables is found to be numerically much smaller than that due to the 
leading PS ($\pi$-like) current. However, these currents give non-negligible 
corrections to isoscalar observables, such as the deuteron magnetic moment 
and B-structure function~\cite{SCH91}. 
Finally it is worth enphasizing that, while the construction
in Ref.~\cite{RIS85} is not 
unique, it has nevertheless been shown to provide, at low and moderate 
values of momentum transfer, satisfactory description of most observables 
where the isovector two-body currents play a large (if not dominant) role, 
such as the deuteron threshold electrodisintegration~\cite{SCH91}, 
the neutron and proton radiative captures on protons and deuterons 
at low energies~\cite{VIV96,SCH91}, and 
the magnetic moments and form factors of the trinucleons (as shown below).

\subsubsection{Two-body charge operators}
\label{subsubsec:2bchop}
While the MI two-body currents are linked to the form of 
nucleon-nucleon interaction 
via the continuity equation, the most important two-body charge operators 
are model dependent and may be viewed as relativistic corrections. 
They fall into two classes. The first class includes those effective 
operators that represent non-nucleonic degrees of freedom, 
such as nucleon-antinucleon pairs or nucleon-resonances, and which arise when 
these degrees of freedom are eliminated from the state vector. To the 
second class belong those dynamical exchange charge effects that would 
appear even in a description explicitly including non-nucleonic excitations 
in the state vector, such as the $\rho\pi\gamma$ and $\omega\pi\gamma$ 
transition couplings. The proper forms of the former operators depend on 
the method of eliminating the non-nucleonic degrees of 
freedom~\cite{FRI77,COO86,COE94}. 
There are nevertheless rather clear indications for the relevance of 
two-body charge operators from the failure of calculations based on 
the one-body operator in Eq.~(\ref{eq:rho1b}) in predicting the charge 
form factors of the three- and four-nucleon systems~\cite{SCH90}, 
and deuteron A-structure function and tensor polarization 
observable~\cite{SCH91,SCHtbp}.

The two-body model used in the present work consists of the $\pi$-, 
$\rho$- and $\omega$-meson exchange charge operators, as well as of the 
$\rho\pi\gamma$ and $\omega\pi\gamma$ charge transition couplings. 
The former are derived by considering the low-energy limit of the 
relativistic Born diagrams associated with the virtual meson 
photoproduction amplitude. The $\rho\pi\gamma$ and $\omega\pi\gamma$ 
operators are the leading corrections obtained in a non-relativistic 
reduction of the corresponding Feynman diagrams with transition couplings, 
for example 
$\langle\pi^{a}(k)\,|\,j_{\mu}(0)\,|\,\rho^{b}(p,\epsilon)\rangle\,=\,
-\left[G_{\rho\pi\gamma}(q_{\mu}^{2})/m_{\rho}\right]\delta_{ab}\epsilon_
{\mu\nu\sigma\tau}\,p^{\nu}k^{\sigma}\epsilon^{\tau}$, $\epsilon$ 
being the polarization vector of the $\rho$-meson. Coupling constants 
and cutoff parameters are given in the previous subsection. 
Explicit expression for all these operators can be found in 
Ref.~\cite{SCH90}. Here we only 
note that: i) the $\pi$- and $\rho$- meson exchange charge operators, 
the former of which gives by far the dominant contribution, are 
constructed using the PS ($\pi$-like) and V ($\rho$-like) components 
projected out of the isospin-dependent spin-spin and tensor terms of 
the interaction~\cite{SCH90}, thus reducing their model dependence. The 
resulting two-body operators are denoted as PS and V, and are here 
obtained from the charge-independent part of the AV18.
ii) In the pion (as well as vector meson) 
charge operators there are additional contributions due to the energy 
dependence of the pion propagator and direct coupling of the photon to 
the exchanged pion ($\rho$-meson). However, these operators give rise 
to non-local isovector contributions which are expected to provide only 
small corrections to the leading local terms. For example these operators 
would only contribute to the isovector combination of the $^{3}$He and 
$^{3}$H charge form factors, which is anyway a factor of three smaller 
than the isoscalar. Thus they are neglected in the present work.

\subsection{The three-body exchange current associated with 
$S$-wave pion rescattering}
\label{subsec:WTcurr}

The isospin odd ``large'' component of the $S$-wave pion-nucleon
($\pi N$) scattering amplitude at low energy and momentum
transfer may be described by the effective 
interaction~\cite{WEI67}:
\begin{equation} 
{\cal L}_{\pi\pi N N}=-{1\over 4 f_\pi^2}\overline{
\psi}\gamma^\mu \bbox{ \tau}\cdot\psi\bbox{\phi}
\times\partial_\mu\bbox{\phi} \ .
\label{eq:WT1}
\end{equation}
Here $\bbox{\phi}$ is the isovector pion field and $f_\pi$ the pion
decay constant ($\simeq $ 93 MeV). This effective Lagrangian
implies the ``Weinberg-Tomozawa'' relation for the isospin odd
combination of the $\pi N$ $S$-wave scattering lengths $a_1,
a_3$:
\begin{equation}
\lambda_2={1\over 6}\left(1+{m_\pi\over m}\right)(a_1 -a_3)=
{1\over 16\pi}\left({m_\pi\over f_\pi}\right)^2 \ ,
\label{eq:WT2}
\end{equation}
which agrees well with the experimental scattering length values.
Combined with the pseudovector $\pi NN$ interaction
\begin{equation}
{\cal L}_{\pi NN}=-{f_{\pi NN}\over m_\pi}\overline{\psi}
\gamma^5\gamma^\mu\bbox{\tau}\psi\cdot\partial_\mu\bbox{\phi} \ ,
\label{eq:WT3}
\end{equation}
where $f_{\pi NN}\simeq 1$, this interaction gives rise to the
three-body interaction:
\begin{eqnarray}
V_S&=&-{1\over 4 m}{1\over f_\pi^2}\left({f_{\pi NN}\over m_\pi}
\right)^2
\sum_{ijk\,{\rm cyclic}}
\bbox{\tau}_i\cdot\bbox{\tau}_j\times\bbox{\tau}_k
{\bbox{ \sigma}_i\cdot {\bf k}_i \bbox{\sigma}_k\cdot{\bf k}_k
\over D_i D_k} \nonumber \\
&&
\Bigg\{\bbox{\sigma}_j\cdot{\bf k}_i\times{\bf k}_k + {{\rm i}\over
2}\left[{\bf k}_i\cdot[({\bf p}_i+{\bf p}_{i}^{'})-({\bf p}_j+
{\bf p}_{j}^{'})]-{\bf k}_k\cdot[({\bf p}_k+{\bf p}_{k}^{'})-({\bf p}_j+
{\bf p}_{j}^{'})]\right]\Bigg\} \ .
\label{eq:WT4}
\end{eqnarray}
The momentum vectors
are defined so that ${\bf k}_i$ denotes the fractional 
momentum transfer to nucleon $i$. The denominator factors $D_i$
are defined as
\begin{equation}
D_i={\bf k}_i^2 + m_\pi^2 \ .
\label{eq:WT5}
\end{equation}

The derivative couplings in the Lagrangians~(\ref{eq:WT1}) 
and~(\ref{eq:WT4})
lead to electromagnetic contact terms. These may be
constructed by minimal substitution, and are found to have the
expressions
\begin{equation} 
{\cal L}_{\pi\pi\gamma N N}=-{1\over 4 f_\pi^2}
\overline{\psi}\gamma^\mu A_\mu[\phi_z(\bbox{\tau}\cdot\bbox{ \phi})-
\tau_z\bbox{\phi}^2]\psi \ ,
\label{eq:WT6}
\end{equation}
\begin{equation}
{\cal L}_{\pi\gamma N N}=-{f_{\pi NN}\over m_\pi}
\overline{\psi}\gamma^5\gamma^\mu A_\mu(\bbox{\tau}\times\bbox{\phi})_z\psi \ ,
\label{eq:WT7}
\end{equation}
respectively. When complemented with the electromagnetic coupling of
the pion,
\begin{equation}
{\cal L}_{\pi\pi \gamma}=-A_\mu(\bbox{\phi}\times
\partial^\mu \bbox{\phi})_z \ ,
\label{eq:WT8}
\end{equation}
these contact terms give rise to the following set of three-nucleon
exchange current operators: (a) a contact current at the $S$-wave
rescattering vertex, (b) two contact currents at the two
accompanying pseudovector $\pi NN$ vertices and (c) two pion
current terms. The explicit expressions for these are in the
corresponding order:
\begin{eqnarray}
{\bf j}_{ijk}^{a}({\bf q})&=&{{\rm i}\over 8 m}{1\over f_\pi^2}
\left({f_{\pi NN}\over m_\pi}\right)^2
[\bbox{\tau}_k\times(\bbox{\tau}_j\times\bbox{\tau}_i)
+\bbox{\tau}_i\times(\bbox{\tau}_j\times\bbox{\tau}_k)]_z \nonumber \\ 
&&
{(\bbox{\sigma}_i\cdot {\bf k}_i)
(\bbox{\sigma}_k\cdot {\bf k}_k)
\over D_i D_k}
[\bbox{\sigma}_j\times({\bf q}-{\bf k}_i-{\bf k}_k)
-{\rm i}({\bf p}_j+{\bf p}_j^{'})] \ , \label{eq:WT9}
\end{eqnarray}
\begin{eqnarray}
{\bf j}_{ijk}^{b}({\bf q})&=&{{\rm i}\over 4m}{1\over f_\pi^2}
\left({f_{\pi NN}\over m_\pi}\right)^2
[\bbox{\tau}_i\times(\bbox{\tau}_j\times\bbox{\tau}_k)]_z
{\bbox{\sigma}_{i}(\bbox{\sigma}_k\cdot{\bf k}_k)\over
D_k D_i{'}}\Bigg\{
[{\bbox{\sigma}}_j\cdot({\bf k}_i-{\bf q})\times
{\bf k}_k]  \Bigg. \nonumber \\
&+&\Bigg. {{\rm i}\over 2}\left[{\bf k}_i\cdot
[({\bf p}_i+{\bf p}_{i}^{'})-({\bf p}_j+
{\bf p}_{j}^{'})]-{\bf k}_k\cdot[({\bf p}_k+{\bf p}_{k}^{'})-
({\bf p}_j+{\bf p}_{j}^{'})] \right. \Bigg. \nonumber \\
&-&\Bigg. \left.  
2 m\,\omega+{\bf q}\cdot({\bf p}_j+{\bf p}_{j}^{'})\right]
\Bigg\} + (i\rightleftharpoons k) \ , 
\label{eq:WT10}
\end{eqnarray}
\begin{eqnarray}
{\bf j}_{ijk}^{c}({\bf q})&=&-{{\rm i}\over 4m}{1\over f_{\pi}^2}
\left({f_{\pi NN}\over m_\pi}\right)^2
[\bbox{\tau}_i\times(\bbox{\tau}_j\times\bbox{\tau}_k)]_z 
{({\bbox{\sigma}}_i\cdot{\bf k}_i)({\bbox{\sigma}}_k\cdot{\bf k}_k)
\over D_i D_k}{2{\bf k}_i-{\bf q}\over D_i{'}} \nonumber \\
&&\Bigg\{[{\bbox{\sigma}}_j\cdot({\bf k}_{i}-{\bf q})\times{\bf k}_k]
+{{\rm i}\over 2}\left[{\bf k}_i\cdot
[({\bf p}_i+{\bf p}_{i}^{'})-({\bf p}_j+
{\bf p}_{j}^{'})]\right. \Bigg. \nonumber \\
&-&\Bigg. \left.{\bf k}_k\cdot[({\bf p}_k+{\bf p}_{k}^{'})-
({\bf p}_j+{\bf p}_{j}^{'})] 
- 2 m\,\omega+{\bf q}\cdot({\bf p}_j+{\bf p}_{j}^{'})\right]
\Bigg\} + (i\rightleftharpoons k) \ .
\label{eq:WT11}
\end{eqnarray}
In these exchange current operators the fractions of the
total momentum transfer ${\bf q}$ imparted to the three
nucleons are denoted ${\bf k}_i$ respectively so that
${\bf q}={\bf k}_1+{\bf k}_2+{\bf k}_3$. The denominator
factors $D_i$ are defined in~(\ref{eq:WT5}), while the denominator
factors $D_i{'}$ are defined as
\begin{equation}
D_i{'}=({\bf q}-{\bf k}_i)^2+m_\pi^2 \ .
\label{eq:WT12}
\end{equation}
The combined three-nucleon exchange current operator
${\bf j}^a+{\bf j}^b+{\bf j}^c$ satifies the continuity
equation with the three-nucleon interaction 
$V_S$~(\ref{eq:WT4}),
as may be verified by comparing the product
${\bf q}\cdot {\bf j}$ with the commutator of $V_S$
and the single nucleon charge operator.
These two-pion exchange three-nucleon 
currents will be labelled as $\pi\pi_S$ below.

Note that the three-nucleon interaction~(\ref{eq:WT4})
is not contained in the Urbana IX three-nucleon
interaction model~\cite{PUD95}, the main part
of which takes into account exchanges that
involve excitation of intermediate $\Delta$-isobar 
resonances, which are treated explicitly below. 
It should however
be included in any complete three-nucleon
interaction model, as it is implied by 
effective chiral Lagrangian models
for the pion-nucleon system. It is included in
three-nucleon interaction models that are based
on pion exchange and rescattering described
by current algebra or chiral Lagrangians
(cf in the ``d'' term of the three-nucleon
interaction in Ref.~\cite{COO93}).

\subsection{Elastic form factors of $^{3}$H and $^{3}$He}
\label{subsec:eff3}
In this section we present results for the magnetic moments, 
charge and magnetic form factors of $^{3}$H and $^{3}$He. 
The nuclear ground states are described by the PHH wave function obtained 
from the AV18/UIX Hamiltonian model. 

A convenient espression to calculate the magnetic 
form factors of a $J$=1/2 nucleus, such as the $A$=3 systems under 
consideration here, is obtained by orienting the coordinate system 
so that the spin-quantization axis (the $z$-axis) lies along the momentum 
transfer ${\bf q}$. It is then found that
\begin{equation}
F_{M}(q)=\frac{2 m}{\mu}\frac{1}{q}\langle\Psi_{+}\,|\,j_{x}
(q \hat{\bf z})\,|\,\Psi_{-}\rangle \ , 
\label{eq:mff} 
\end{equation}
where $\mu$ is the nuclear magnetic moment in terms of $\mu_{N}$ 
(the nuclear magneton), $\Psi_{+/-}$ are the normalized ground state 
wave functions with $J_{z}=\pm$1/2, respectively, and $j_{x}(q \hat{\bf z})$ 
is the $x$-component of the current operator. Note that $F_{M}(0)=$1.
The charge form factor is easily obtained from 
\begin{equation}
F_{C}(q)=\frac{1}{Z}\langle\Psi_{+}\,|\,\rho(q \hat{\bf z})\,|\,
\Psi_{+}\rangle \ ,
\label{eq:cff}
\end{equation}
with $F_{C}(0)=1$.

The matrix elements~(\ref{eq:mff}) and~(\ref{eq:cff}) are evaluated 
with Monte Carlo methods. The wave function is written as a vector 
in the spin-isospin space of the three nucleons for any given spatial 
configuration ${\bf R}\equiv({\bf r}_{1}, {\bf r}_{2}, {\bf r}_{3})$. 
For the given ${\bf R}$, the state vectors $j_{x}(q \hat{\bf z})\,|\,
\Psi_{-}\rangle$ and $\rho(q \hat{\bf z})\,|\,\Psi_{+}\rangle$ are 
calculated by performing exactly the spin-isospin algebra with the 
techniques described in Refs.~\cite{SCH89,SCH90}. The spatial integrations 
are carried out by sampling the ${\bf R}$-configurations according to the 
Metropolis $et$ $al$. algorithm~\cite{MET53}. 
Typically, 400,000 configurations are 
enough, in the form factor calculations reported here, to achieve a relative 
error of few \% at low and moderate values of momentum transfer 
$q$ ($q\leq$ 5 fm$^{-1}$), increasing to $\sim$30\% at the highest 
$q$-values.

\subsubsection{The magnetic form factors}
\label{subsubsec:mff}
The current operator includes, in addition to the one-body current in 
Eq.~(\ref{eq:j1bNN}), the MI two-body currents obtained from the 
charge-independent part of the AV18 interaction (denoted as PS or 
$\pi$-like, V or $\rho$-like, SO, LL and SO2), the MD $\rho\pi\gamma$ 
and $\omega\pi\gamma$ two-body currents, and finally the local terms 
of the three-body current associated with the $S$-wave two-pion
exchange three-nucleon interaction~(\ref{eq:WT4})
described in the previous section.

Because of destructive interference in the matrix element for the magnetic 
dipole transition between the $S$- and $D$-state components of the wave 
function, the one-body predictions for the $^{3}$H and $^{3}$He magnetic 
form factors (MFF) have distinct minima at around $\sim$3.5 fm$^{-1}$ and 
$\sim$2.5 fm$^{-1}$, respectively, in disagreement with the experimental 
data~\cite{COL65,MCC77,SZA77,ARN78,CAV82,DUN83,OTT85,JUS85,BEC87,AMR94}, 
as shown in Figs.~\ref{fig:mff3hNN} and~\ref{fig:mff3heNN}. The situation 
is closely related to that of the backward cross section for 
electrodisintegration of the deuteron, which is in fact dominated by 
two-body current contributions for values of momentum transfer above 
$\sim$2.5 fm$^{-1}$~\cite{HOC73}.

Inclusion of the contributions from the two- and $\pi\pi_S$ 
three-body currents 
shifts the zeros in the calculated MFF to higher $q$-values. While the 
experimental $^{3}$H MFF is in good agreement with theory over a wide 
range of momentum transfers, there is a significant discrepancy 
between the measured and calculated values of the $^{3}$He MFF in the region 
of diffraction minimum. This discrepancy persists even when different 
parametrizations of the nucleon electromagnetic form factors are used. 
This is evident from 
Figs.~\ref{fig:mff3hNN} and~\ref{fig:mff3heNN} where the total results 
obtained with the Gari-Kr\"{u}mpelmann (GK) parametrization~\cite{GAR86} 
of the nucleon electromagnetic form factors are shown.

It is useful to define the quantities
\begin{equation}
F_{M}^{S,V}(q)=\frac{1}{2}\left[\mu(^{3}{\rm He})F_{M}(q;^{3}{\rm He})
\pm\mu(^{3}{\rm H})F_{M}(q;^{3}{\rm H})\right] \ .
\label{eq:isvscmff}
\end{equation}
If the $^{3}$H and $^{3}$He ground states were pure $T$=1/2 states, 
then $F_{M}^{S}$ and $F_{M}^{V}$ linear combinations of the three-nucleon 
MFF would be only influenced by, respectively, the isoscalar ($S$) and 
isovector ($V$) parts of the current operator. However, small isospin 
admixtures with $T>$1/2, induced by electromagnetic, CSB and CIB terms 
present in the AV18 interaction, are included in the present wave functions. 
As a consequence, purely isoscalar (isovector) current operators give small, 
otherwise vanishing, contributions to the $F_{M}^{V}$ ($F_{M}^{S}$) MFF.

The contributions of the individual components of the two- and 
three-nucleon ($\pi\pi_S$ term) currents to the $F_{M}^{S}$ and $F_{M}^{V}$ 
combinations are shown in Figs.~\ref{fig:mffis_cNN} and~\ref{fig:mffiv_cNN}. 
In the diffraction region the PS ($\pi$-like) isovector current gives the 
dominant contribution to $F_{M}^{V}$, while the contributions 
from remaining currents are significantly smaller, about one order of 
magnitude or more. The three-nucleon current 
($\pi\pi_S$) associated with the $S$-wave $\pi N$ coupling 
is found to give a very small correction.

Among the two-body contributions to $F_{M}^{S}$, the most important is that 
due to the currents from the spin-orbit interactions (SO), the remaining 
operators producing a very small correction. Note that the isovector PS and 
V currents contribute to $F_{M}^{S}$ because of the small isospin-symmetry 
breaking components present in the $^{3}$H and $^{3}$He wave functions 
induced by the AV18 model, as mentioned earlier.

Finally, the cumulative contributions to the $F_{M}^{S}$ and $F_{M}^{V}$ 
combinations are compared with the experimental data~\cite{AMR94} in 
Figs.~\ref{fig:mffisNN} and~\ref{fig:mffivNN}, respectively. The zero in the 
calculated $F_{M}^{V}$ is found to occur at lower $q$-value than 
experimentally observed. As shown in the next section, this discrepancy 
between theory and experiment remains unresolved even when $\Delta$-isobar 
degrees 
of freedom are included in both the nuclear wave functions and currents. 
We will return to this point in the conclusions. Predictions for the 
magnetic moments are given in Tables~\ref{tb:mmomcontr} and~\ref{tb:mmom}, 
while those for the magnetic radii are listed in Table~\ref{tb:mgr}.
These results are discussed in Sec. III-D.
%~\ref{subsec:mmom+mff}.

\subsubsection{The charge form factors}
\label{subsubsec:cff}
The charge operator includes, in addition to the one-body term of 
Eq.~(\ref{eq:rho1b}), the PS or $\pi$-like, V or $\rho$-like, $\omega$, 
$\rho\pi\gamma$ and $\omega\pi\gamma$ two-body operators, discussed 
previously. The calculated $^{3}$H and $^{3}$He charge form factors (CFF) 
are compared with the experimental data
~\cite{COL65,MCC77,SZA77,ARN78,CAV82,DUN83,OTT85,JUS85,BEC87,AMR94} 
in Figs.~\ref{fig:cff3hNN} and
~\ref{fig:cff3heNN}. There is excellent agreement between theory and 
experiment, as is clear from these figures. The important role of the 
two-body contributions above 3 fm$^{-1}$ is also evident. 
The remarkable success of the present picture based on non-relativistic 
wave functions and a charge operator including the leading relativistic 
corrections should be stressed. It suggests, in particular, that the 
present model for the two-body charge operator is better than one 
{\it a priori} should expect. These operators, such as the PS 
charge operator, fall into the class of relativistic corrections. Thus, 
evaluating their matrix elements with non-relativistic wave functions 
represents only the first approximation to a systematic reduction. A 
consistent treatment of these relativistic effects would require, for 
example, inclusion of the boost corrections on the nuclear wave 
functions~\cite{FRI77,COO86,SCH96}. 
Yet, the excellent agreement between the calculated 
and measured CFF suggests that these corrections may be neglegible in 
the $q$-range explored so far.

For completeness, we show 
in Figs.~\ref{fig:cffis_cNN} and~\ref{fig:cffiv_cNN} 
the contributions from the individual components 
of the charge operator to the linear combinations
\begin{equation}
F_{C}^{S,V}(q)=\frac{1}{2}\left[2\,F_{C}(q;^{3}{\rm He})\pm F_{C}
(q;^{3}{\rm H})\right] \ .
\label{eq:isvsccff}
\end{equation}
Note that again, because of isospin-symmetry breaking components present in 
the $^{3}$He and $^{3}$H wave functions, the purely isovector (isoscalar) 
$\omega\pi\gamma$ ($\rho\pi\gamma$) charge operator gives a small, 
otherwise vanishing, correction to the $F_{C}^{S}$ ($F_{C}^{V}$) CFF.

Finally, values for the charge radii of $^{3}$H and $^{3}$He
are listed in Table~\ref{tb:chr}. The results 
including the contributions associated 
with the two-body charge operators are found to be in good agreement 
with experimental data.

\section{Beyond nucleons only}
\label{sec:ND}
The simplest picture views the nucleus as being made up of nucleons, and 
assumes that all other sub-nucleonic degrees of freedom may be eliminated 
in favor of effective many-body operators acting on the nucleons' 
coordinates. The validity of such a description is based on the success 
it has achieved in the quantitative prediction of many nuclear 
observables~\cite{CAR98}. However, it is interesting to consider 
corrections to this picture by including the degrees of freedom associated 
with nuclear resonances as additional constituents of the nucleus. 
When treating phenomena which do not involve explicitely meson production, 
it is reasonable to expect that the lowest excitation of the nucleon, 
the $\Delta$-isobar, plays a leading role.

In such an approach, the $A$=3 nuclear wave function is written as
\begin{equation}
\Psi_{N+\Delta} = \Psi(NNN) + \Psi^{(1)}(NN\Delta) + \Psi^{(2)}
(N\Delta\Delta) + \Psi^{(3)}(\Delta\Delta\Delta) \ ,
\label{eq:psiND}
\end{equation}
where $\Psi$ is that part of the total wave function consisting only of 
nucleons; the term $\Psi^{(1)}$ is the component in which a single nucleon 
has been converted into a $\Delta$-isobar, and so on. The nuclear 
two-body interaction is taken as
\begin{equation}
v_{ij}=\sum_{B_{i},B_{j}=N,\Delta}\sum_{B_{i}^{'},B_{j}^{'}=N,\Delta} v_{ij}
(B_{i}B_{j}\rightarrow B_{i}^{'}B_{j}^{'}) \ ,
\label{eq:tbi}
\end{equation}
where transition interactions such as $v_{ij}(NN\rightarrow N\Delta)$, 
$v_{ij}(NN\rightarrow \Delta\Delta)$, etc. are responsible for generating 
$\Delta$-isobar admixtures in the wave function. The long-range part of 
$v_{ij}$ is due to pion-exchange, while its short- and intermediate-range 
parts, influenced by more complex dynamics, is constrained by fitting $NN$ 
scattering data at lab energy $\leq$ 400 MeV and deuteron 
properties~\cite{WIR84}.

Once the $NN$, $N\Delta$ and $\Delta\Delta$ interactions have been 
determined, the problem is reduced to solving the $N$-$\Delta$ 
coupled-channel Schr\"{o}dinger equation. In principle, for the 
$A$=3 systems Faddeev and hyperspherical-harmonics techniques can be 
used (and, indeed, Faddeev methods have been used in the past
~\cite{HAJ83,PIC92}) to this end, although the large number of $N$-$\Delta$ 
channels involved makes the practical implementation of these methods 
difficult. A somewhat simpler approach consists of a generalization 
of the correlation operator technique~\cite{LOM81}, which has proven very 
useful in the variational theory of light nuclei, particularly in the 
context of variational Monte Carlo calculations~\cite{PUD97,WIR91}. 
In such an approach, known as the transition-correlation-operator (TCO) 
method~\cite{SCH92}, the nuclear wave function is written as
\begin{equation}
\Psi_{N+\Delta}=\left[{\cal{S}}\prod_{i<j}\left(1\,+\,U^{TR}_{ij}\right)
\right]\,\Psi \ ,
\label{eq:psiNDtco}
\end{equation}
where $\Psi$ is the purely nucleonic component, $\cal{S}$ is the 
symmetrizer and the transition operators $U^{TR}_{ij}$ convert $NN$ pairs 
into $N\Delta$ and $\Delta\Delta$ pairs. In the present study the $\Psi$ 
is taken from PHH solutions of the AV18/UIX Hamiltonian with nucleons only 
interactions~\cite{SCH92}, while the $U^{TR}_{ij}$ is obtained from two-body 
bound and low-energy scattering state solutions of the full $N$-$\Delta$ 
coupled-channel problem. This aspect of the present calculations is 
reviewed briefly in the next subsection.

\subsection{Wave functions with $\Delta$-admixtures}
\label{subsec:deltawf}
The transition correlation operator (TCO) method~\cite{SCH92} 
consists in approximating the $\Psi_{N+\Delta}$ as in 
Eq.~(\ref{eq:psiNDtco}), with the transition operators $U^{TR}_{ij}$ 
defined as
\begin{equation}
U_{ij}^{TR}\,=\,U_{ij}^{N\Delta}\,+\,U_{ij}^{\Delta N}\,+\,U_{ij}^
{\Delta\Delta} \ , 
\label{eq:tcoop}
\end{equation}
\begin{eqnarray}
U_{ij}^{N\Delta}&=&\left[u^{\sigma\tau II}(r_{ij})\bbox{\sigma}_{i}\cdot
{\bf S}_{j}\,+\,u^{t\tau II}(r_{ij})S_{ij}^{II}\right]\,\bbox{\tau}_{i}
\cdot{\bf T}_{j} \ , \label{eq:tcops1}\\
U_{ij}^{\Delta\Delta}&=&\left[u^{\sigma\tau III}(r_{ij}){\bf S}_{i}\cdot
{\bf S}_{j}\,+\,u^{t\tau III}(r_{ij})S_{ij}^{III}\right]\,{\bf T}_{i}\cdot
{\bf T}_{j} \ .
\label{eq:tcops2}
\end{eqnarray}
Here, ${\bf S}_{i}$ and ${\bf T}_{i}$ are spin- and isospin-transition 
operators which convert nucleon $i$ into a $\Delta$-isobar; $ S_{ij}^{II}$ 
and $S_{ij}^{III}$ are tensor operators in which, respectively, the Pauli 
spin operators of either particle $i$ or $j$, and both particles $i$ and $j$ 
are replaced by corresponding spin-transition operators. The $U_{ij}^{TR}$ 
vanishes in the limit of large interparticle separations, since no 
$\Delta$-components can exist asymptotically.

The transition operator $U_{ij}^{TR}$ and nucleonic wave function $\Psi$ 
in Eq.~(\ref{eq:psiNDtco}) could be determined variationally by using an 
interaction of the form given in Eq.~(\ref{eq:tbi}), that contains both 
$N$ and $\Delta$ degrees of freedom, such as the Argonne $v_{28Q}$ (AV28Q) 
model~\cite{WIR84,WIRpc}, and by minimizing the ground-state energy of 
each given nucleus. Instead, we use transition correlation functions 
$u^{\sigma\tau II}(r)$, etc. (shown in Fig.~\ref{fig:tcf}) that approximately 
reproduce two-body bound- and low-energy scattering-state wave functions 
for the AV28Q model, and take the PHH wave function obtained in 
Sec. II-A 
%~\ref{subsec:ham+wf} 
as the $\Psi$ in Eq.~(\ref{eq:psiNDtco}). 
The validity of such an approximation has been discussed at length in 
the original reference~\cite{SCH92}. Here, we only note that i) since 
the correlation functions $u^{\sigma\tau II}(r)$, etc. are short-ranged, 
they are expected to have a rather weak dependence on $A$; ii) it is 
important the $\Psi$ used in Eq.~(\ref{eq:psiNDtco}), obtained from a 
$v_{ij}(NN\rightarrow NN)$ interaction phase-equivalent to the full 
$v_{ij}$ of Eq.~(\ref{eq:tbi}), be proportional to that projected out from 
the $\Psi_{N+\Delta}$ wave function for the $v_{ij}$ interaction. This has 
been explicitly verified by direct calculation 
in the two-body problem~\cite{SCH92}.

In the TCO scheme, the perturbation theory description of $\Delta$-admixtures 
is equivalent to the replacements:
\begin{eqnarray}
U_{ij}^{N\Delta,{\rm PT}}&=&\frac{v_{ij}(NN\rightarrow N\Delta)}
{m-m_{\Delta}} \ , \label{eq:UNDpt} \\
U_{ij}^{\Delta\Delta,{\rm PT}}&=&\frac{v_{ij}(NN\rightarrow \Delta\Delta)}
{2(m-m_{\Delta})} \ , \label{eq:UDDpt} 
\end{eqnarray}
where the kinetic energy contributions in the denominators of 
Eqs.~(\ref{eq:UNDpt}) and~(\ref{eq:UDDpt}) have been neglected 
(static $\Delta$ approximation). 
The transition interactions $v_{ij}(NN\rightarrow N\Delta)$ 
and $v_{ij}(NN\rightarrow \Delta\Delta)$ have the same operator structure 
as $U_{ij}^{N\Delta}$ and $U_{ij}^{\Delta\Delta}$ of Eqs.~(\ref{eq:tcops1}) 
and~(\ref{eq:tcops2}), but with 
the $u^{\sigma\tau\alpha}(r)$ and $u^{t\tau\alpha}(r)$ functions 
replaced by, respectively,
\begin{eqnarray}
v^{\sigma\tau\alpha}(r)&=&
\frac{(ff)_{\alpha}}{4\pi}\frac{m_{\pi}}{3}\frac{e^{-x}}{x}\,C(x) \ ,
\label{eq:uvst} \\
v^{t\tau\alpha}(r)&=&
\frac{(ff)_{\alpha}}{4\pi}\frac{m_{\pi}}{3}\left(1+\frac{3}{x}+\frac{3}{x^2}
\right)\frac{e^{-x}}{x}\,C^{2}(x) \ .
\label{eq:uvtt} 
\end{eqnarray}
Here $\alpha$ = II, III, $x\equiv m_{\pi}r$, $(ff)_{\alpha}=f_{\pi NN}
f_{\pi N \Delta}$, $f_{\pi N \Delta}f_{\pi N \Delta}$ for $\alpha$ = II, III, 
respectively, and the cutoff function $C(x)\,=\,1-e^{-\lambda x^{2}}$, 
$\lambda$ = 4.09 in the AV28Q model~\cite{WIRpc}. Note that 
in Fig.~\ref{fig:tcf} $u^{q \rm{II, PT}}(r)=v^{q \rm{II}}(r)/(m-m_{\Delta})$ 
and $u^{q \rm{III, PT}}(r)=v^{q \rm{III}}(r)/[2(m-m_{\Delta})]$, 
with $q$ = $\sigma\tau$, $t\tau$.

The perturbative treatment has been often (in fact, almost exclusively) 
used in the literature to estimate the effect of $\Delta$ degrees of freedom 
on electroweak observables. However, it may lead to a substantial 
overprediction of their importance~\cite{VIV96,SCH92}, 
since it produces $N\Delta$ and $\Delta\Delta$ wave functions 
which are too large at short distance, see Fig.~\ref{fig:tcf}.

\subsection{$N\Delta$-transition and $\Delta$ currents}
\label{subsec:ND+Dcurr}
The nuclear electromagnetic current is expanded into a sum of many-body 
terms that operate on the nucleon and $\Delta$-isobar degrees of freedom. 
The nucleonic component of this current operator has been discussed in the 
previous section. Here, we only discuss its $\Delta$ components.

\subsubsection{$N\Delta$-transition and $\Delta$ one-body currents}
\label{subsubsec:1bND+Dcurr}
The one-body current is written as 
\begin{equation}
{\bf j}^{(1)}_{i}({\bf q})=\sum_{B,B^{'}=N,\Delta}{\bf j}_{i}({\bf q};
B\rightarrow B^{'}) \ ,
\label{eq:1bterm}
\end{equation}
where ${\bf j}_{i}({\bf q};N\rightarrow N)$ is the nucleonic current 
component given in Eq.~(\ref{eq:j1bNN}) and
\begin{eqnarray}
{\bf j}_{i}({\bf q};N\rightarrow \Delta)&=&-\frac{\rm i}{2 m}
G_{\gamma N\Delta}(q_{\mu}^{2}) e^{{\rm i}{\bf q}\cdot{\bf r}_{i}} 
{\bf q}\times{\bf S}_{i} T_{z,i} \ , \label{eq:j1bND} \\
{\bf j}_{i}({\bf q};\Delta\rightarrow \Delta)&=&-\frac{\rm i}{24 m}
G_{\gamma\Delta\Delta}(q_{\mu}^{2}) e^{{\rm i}{\bf q}\cdot{\bf r}_{i}} 
{\bf q}\times\bbox{\Sigma}_{i} (1+\Theta_{z,i}) \ . \label{eq:j1bDD}
\end{eqnarray}
Here $\bbox{\Sigma}$ ($\bbox{\Theta}$) is the Pauli operator for the 
$\Delta$ spin 3/2 (isospin 3/2), and the expression for ${\bf j}_{i}({\bf q};
\Delta\rightarrow N)$ is obtained from that for ${\bf j}_{i}({\bf q};
N\rightarrow \Delta)$ by replacing the transition spin and isospin 
operators by their hermitian conjugates. The $N\Delta$-transition and 
$\Delta$ electromagnetic form factors, respectively $G_{\gamma N\Delta}$ 
and $G_{\gamma\Delta\Delta}$, are parametrized as
\begin{eqnarray}
G_{\gamma N\Delta}(q_{\mu}^{2})&=&\frac{\mu_{\gamma N\Delta}}
{\left(1+q_{\mu}^{2}/\Lambda^{2}_{N\Delta,1}\right)^{2}\sqrt{1+q_{\mu}^{2}/
\Lambda^{2}_{N\Delta,2}}} \ , \label{eq:gff1}\\
G_{\gamma \Delta\Delta}(q_{\mu}^{2})&=&\frac{\mu_{\gamma \Delta\Delta}}
{\left(1+q_{\mu}^{2}/\Lambda^{2}_{\Delta\Delta}\right)^{2}} \ .  
\label{eq:gff2}
\end{eqnarray}
Here the $N\Delta$-transition magnetic moment $\mu_{\gamma N\Delta}$ is 
taken equal to 3 $\mu_{N}$, as obtained from an analysis of $\gamma N$ 
data in the $\Delta$-resonance region~\cite{CAR86}; this analysis also 
gives $\Lambda_{N\Delta,1}\,=$ 0.84 GeV and $\Lambda_{N\Delta,2}\,=$ 1.2 GeV. 
The value used for the $\Delta$ magnetic moment $\mu_{\gamma\Delta\Delta}$ 
is 4.35 $\mu_{N}$ by averaging results of a soft-photon analysis of 
pion-proton bremsstrahlung data near the 
$\Delta^{++}$ resonance~\cite{LIN91}, 
and $\Lambda_{\Delta\Delta}\,=$ 0.84 GeV as in the dipole 
parametrization of the nucleon form factor. In principle, $N$ to $\Delta$ 
excitation can also occur via an electric quadrupole transition. 
Its contribution, however, has been ignored, since the associated pion 
photoproduction amplitude is found to be experimentally small at 
resonance~\cite{ERI88}. Also neglected is the $\Delta$ convection current.

\subsubsection{$N\Delta$-transition two-body currents}
\label{subsubsec:2bNDcurr}
The two-body term is written as
\begin{equation}
{\bf j}_{ij}^{(2)}({\bf q})\,=\,{\sum}'_{B_{i}, B_{j}=N, \Delta}{\sum}'
_{B_{i}^{'}, B_{j}^{'}=N, \Delta}\,\, {\bf j}_{ij}({\bf q}; B_{i}B_{j}
\rightarrow B_{i}^{'}B_{j}^{'}) \ , \label{2bterm}
\end{equation}
where the prime over the summation symbols indicates that terms involving 
more than a single $\Delta$ have been neglected in the present study. The 
$NN\rightarrow NN$ two-body terms have already been discussed. The 
two-body terms involving at most a single $\Delta$ are illustrated in 
Fig.~\ref{fig:cur2Delta}, and are explicitly given by 
\begin{eqnarray}
{\bf j}_{ij}({\bf q}; NN\rightarrow N\Delta)&=&(\bbox{\tau}_{i}\times
{\bf T}_{j})_{z} \Bigg[\left[\bbox{\sigma}_{i}({\bf S}_{j}\cdot\hat
{\bf r}_{ij}) e^{{\rm i}{\bf q}\cdot{\bf r}_{i}} + (\bbox{\sigma}_{i}\cdot
\hat{\bf r}_{ij}){\bf S}_{j} e^{{\rm i}{\bf q}\cdot{\bf r}_{j}}\right] 
h(r_{ij}) \Bigg. \nonumber \\
&+&\Bigg. e^{{\rm i}{\bf q}\cdot{\bf R}_{ij}}(\bbox{\sigma}_{i}\cdot\bbox
{\nabla}_{i})({\bf S}_{j}\cdot\bbox{\nabla}_{j}) \hat{\bf r}_{ij}
{\overline h}(r_{ij})\Bigg] \ ,
\label{eq:2bND}
\end{eqnarray}
where ${\bf r}_{ij}={\bf r}_{i} - {\bf r}_{j}$, ${\bf R}_{ij}=
({\bf r}_{i}+{\bf r}_{j})/2$, and the functions $h(r)$ and 
${\overline h}(r)$ are defined as, respectively,
\begin{eqnarray}
h(r)&\equiv&-\left(\frac{f_{\pi NN}f_{\pi N\Delta}}{4\pi}\right)
\frac{1}{x^2}(1+x)e^{-x} \ , \label{eq:hr} \\
\bar{h}(r)&\equiv&\left(\frac{f_{\pi NN}f_{\pi N\Delta}}{4\pi}\right)
\frac{1}{m_{\pi}^{2}}\int_{-\frac{1}{2}}^{+\frac{1}{2}}dz\,
e^{-{\rm i}z{\bf q}\cdot{\bf r}}e^{-r L(z)} \ , \label{eq:hbr} 
\end{eqnarray}
with $x=m_{\pi}r$ and $L(z) = [ m_{\pi}^{2} + q^{2}(1/4 - z^{2})]^{1/2}$. 
Terms explicitly proportional to ${\bf q}$ in 
Eq.~(\ref{eq:2bND}) have been dropped, since in applications only 
the transverse components of ${\bf j}({\bf q})$ occur. The three terms 
in Eq.~(\ref{eq:2bND}) are associated with diagrams (a), (b) and (c) 
in Fig.~\ref{fig:cur2Delta}, respectively, and can be obtained from the 
well known expression of the two-body nucleonic currents due to 
pion-exchange by replacing $\bbox{\sigma}_{j}$ and $\bbox{\tau}_{j}$ 
with ${\bf S}_{j}$ and ${\bf T}_{j}$, respectively.

To account for the hadron extended structure, form factors must be 
introduced at the $\pi NN$ and $\pi N\Delta$ vertices. In the case 
of $v_{ij}(NN\rightarrow N\Delta)$ interaction, an $r$-space gaussian 
cutoff has been used. However, for the $j(NN\rightarrow N\Delta)$ above 
it has been found convenient to introduce monopole form factors with 
$\Lambda$=900 MeV in its $p$-space expression. 
This value for $\Lambda$ is consistent with that obtained 
from the tensor component of $v_{ij}(NN\rightarrow N\Delta)$. Finally, 
the expression in Eq.~(\ref{eq:2bND}) is multiplied by the isovector form 
factor $G_{E}^{V}(q_{\mu}^{2})$.

\subsection{Calculation}
\label{subsec:calmff}
Calculation of the magnetic form factors requires evaluation of the 
transition matrix element in Eq.~(\ref{eq:mff}), where the wave functions 
and currents include both nucleonic and $\Delta$-isobar degrees of freedom. 
To evaluate such a matrix element, it is convenient to expand the wave 
function $\Psi_{N+\Delta,J_{z}}$ as
\begin{equation}
\Psi_{N+\Delta,J_{z}} = \Psi_{J_{z}} + \sum_{i<j}U_{ij}^{TR}\Psi_{J_{z}} + 
\ldots \ ,
\label{eq:psind}
\end{equation}
and write the numerator of Eq.~(\ref{eq:mff}), in a schematic notation, as
\begin{equation}
\langle\Psi_{N+\Delta,f}\,|\,j\,|\,\Psi_{N+\Delta,i}\rangle\,=\,\langle
\Psi_{f}\,|\,j(N\,{\rm only})\,|\,\Psi_{i}\rangle 
\,+\,\langle\Psi_{f}\,|\,j(\Delta)\,|\,\Psi_{i}\rangle \ ,
\label{eq:schemj}
\end{equation}
where $j(N \,{\rm only})$ denotes all one- and two-body contributions to 
${\bf j}({\bf q})$ which only involve nucleon degrees of freedom, i.e., 
$j(N \,{\rm only})\,=\,j^{(1)}(N\rightarrow N)\,+\,j^{(2)}(NN\rightarrow NN)$. 
The operator $j(\Delta)$ includes terms involving the $\Delta$-isobar 
degrees of freedom, associated with the explicit $\Delta$ currents $j^{(1)}
(N\rightarrow\Delta)$, $j^{(1)}(\Delta\rightarrow N)$, $j^{(1)}(\Delta
\rightarrow\Delta)$, $j^{(2)}(NN\rightarrow N\Delta)$, etc., and with the 
transition operators $U_{ij}^{TR}$. The operator $j(\Delta)$ is illustrated 
diagrammatically in Figs.~\ref{fig:1bD} and~\ref{fig:2bD}. The terms (a)-(g) 
in Fig.~\ref{fig:1bD} and (a)-(f) in Fig.~\ref{fig:2bD} are two-body current 
operators. The terms (g)-(l) in Fig.~\ref{fig:2bD} are three-body current 
operators, while the terms (h)-(j) in Fig.~\ref{fig:1bD} are to be 
interpreted as renormalization corrections to the ``nucleonic'' matrix 
elements $\langle\Psi_{f}\,|\,j(N {\rm only})\,|\,\Psi_{i}\rangle$, due 
to the presence of $\Delta$-admixtures in the wave functions.

There are, however, additional, connected three-body terms in $j(\Delta)$ 
that are neglected in the present work. A number of these are illustrated 
in Fig.~\ref{fig:3bDnegl}. Their contribution is expected to be significantly 
smaller than that from the terms in Figs.~\ref{fig:1bD} and~\ref{fig:2bD} 
involving transition correlations between two particles only, of the 
type ${U_{ij}^{B B^{'}}}^{\dagger}\,U_{ij}^{B B^{'}}$, but comparable to 
that from the three-body terms in Fig.~\ref{fig:2bD} having ${U_{ij}^
{B B^{'}}}^{\dagger}\,U_{jk}^{B B^{'}}$. These have been found to be very 
small.

The terms in Fig.~\ref{fig:1bD} are expanded as operators acting on the 
nucleons' coordinates. For example, the terms (a) and (e) in 
Fig.~\ref{fig:1bD} have the structure, respectively,
\begin{eqnarray}
({\rm a})&=&j_{i}^{(1)}(\Delta\rightarrow N)\,U_{ij}^{\Delta N} \ , 
\label{eq:aterm} \\
({\rm e})&=&{U_{ij}^{\Delta N}}^{\dagger}\,j^{(1)}_{i}
(\Delta\rightarrow\Delta)\,U_{ij}^{\Delta N} \ , \label{eq:eterm}
\end{eqnarray}
which can be reduced to operators involving only Pauli spin and isospin 
matrices by using the identities
\begin{eqnarray}
{\bf S}^{\dagger}\cdot{\bf A}\,{\bf S}\cdot{\bf B}&=&\frac{2}{3}{\bf A}
\cdot{\bf B}-\frac{\rm i}{3}\bbox{\sigma}\cdot ({\bf A}\times{\bf B}) \ ,
\label{eq:SdagS} \\
{\bf S}^{\dagger}\cdot{\bf A}\,\bbox{\Sigma}\cdot{\bf B}\,{\bf S}
\cdot{\bf C}&=&\frac{5}{3}\,{\rm i}\,{\bf A}\cdot({\bf B}\times{\bf C})-
\frac{1}{3}\bbox{\sigma}\cdot{\bf A}\,{\bf B}\cdot{\bf C} \nonumber \\
& & -\frac{1}{3}{\bf A}\cdot {\bf B}\,{\bf C}\cdot\bbox{\sigma}+\frac{4}{3}
{\bf A}\cdot({\bf B}\cdot\bbox{\sigma}){\bf C} \ ,
\label{eq:SdagSigmaS}
\end{eqnarray}
where ${\bf A}$, ${\bf B}$ and ${\bf C}$ are vector operators that commute 
with $\bbox{\sigma}$, but not necessarily among themselves.

While the terms in Fig.~\ref{fig:2bD} could have been reduced in precisely 
the same way, the resulting expressions in terms of $\bbox{\sigma}$ and 
$\bbox{\tau}$ Pauli matrices become too cumbersome. Thus, for these it was 
found to be more convenient to retain the explicit representation of 
${\bf S}$ $({\bf S}^{\dagger})$ as a $4 \times 2\,(2\times 4)$ matrix

$ \hspace{5cm}
  {\bf S}= \left( \begin{array}{cc}
           -\hat{{\bf e}}_{-} & 0  \\
           \sqrt{\frac{2}{3}}\hat{{\bf e}}_{0} & -\frac{1}{\sqrt{3}}
	                     \hat{{\bf e}}_{-} \\
           -\frac{1}{\sqrt{3}}\hat{{\bf e}}_{+}& \sqrt{\frac{2}{3}}
                              \hat{{\bf e}}_{0} \\
	   0  & -\hat{{\bf e}}_{+}
           \end{array} \right) \ , $

\noindent
where $\hat{{\bf e}}_{\pm}=\mp(\hat{\bf x}\pm {\rm i}\hat{\bf y})/{\sqrt{2}}$, 
$\hat{\bf e}_{0}=\hat{\bf z}$, and $\hat{\bf e}_{\mu}^{*}=
(-)^{\mu}\hat{\bf e}_{-\mu}$ and derive the result of terms such as 
(a)+(c)+(e)=${U_{ij}^{N\Delta}}^{\dagger}\,j^{(2)}_{ij}
(NN\rightarrow N\Delta)$ on the state $|\Psi\rangle$ by first operating 
with $j^{(2)}$ and then with ${U^{N\Delta}}^{\dagger}$. The Monte Carlo 
evaluation of the matrix element is then performed with methods similar 
to those sketched in Sec. II-D.
%~\ref{subsec:eff3}

The normalization of the wave function is given by
\begin{eqnarray}
\langle \Psi_{N+\Delta,J_{z}}\,|\,\Psi_{N+\Delta,J_{z}}\rangle&=&
\langle \Psi_{J_{z}}\,|\,1\,+\,\sum_{i<j}[2\,{U_{ij}^{\Delta N}}^{\dagger}
U_{ij}^{\Delta N}\,+\,{U_{ij}^{\Delta \Delta}}^{\dagger}U_{ij}^{\Delta 
\Delta}]
\,|\,\Psi_{J_{z}}\rangle \nonumber \\
&+&\,(\,{\rm three\!-\!body\, terms})
\label{eq:normal}
\end{eqnarray}
and the three-body terms have been neglected consistently with the 
approximation introduced in Eq.~(\ref{eq:schemj}), as discussed above.

Perturbation theory (PT) estimates of the importance of $\Delta$-isobar 
degrees of freedom in photo- and electro-nuclear observables typically 
include only the contribution from single $N\rightleftharpoons\Delta$ 
transitions (namely diagrams (a) and (b) in Fig.~\ref{fig:1bD}) and ignore 
the change in the wave function normalization. In the TCO scheme, these PT 
estimates are obtained by using $U^{B B^{'},PT}$ transition correlation 
defined in Eqs.~(\ref{eq:UNDpt}) and~(\ref{eq:UDDpt})~\cite{SCH92}. In 
particular, the PT expressions for the three-body terms in 
Fig.~\ref{fig:2bD}, diagrams (g)-(h)along with those in which the first 
and third legs are exchanged, can easily be shown to satisfy current 
conservation with the Fujita-Miyazawa two-pion exchange three-nucleon 
interaction (TNI)~\cite{FUJ57} given by
\begin{equation}
V_{ijk}^{FM}(NNN \rightarrow NNN) = v_{jk}(\Delta N\rightarrow NN)\frac{1}
{m-m_{\Delta}}v_{ij}(NN \rightarrow N\Delta) +{\rm h.c.} \  ,
\label{eq:v3bfm}
\end{equation}
where the transition potentials are defined in Eqs.~(\ref{eq:uvst}) and
~(\ref{eq:uvtt}) (here with the cutoff function $C(x)$ set to one). 
Current models of TNI~\cite{PUD95} include the ``long-range'' 
$2\pi$-exchange component above. Indeed, the need of including the 
associated three-body currents provided one of the motivations for 
undertaking the present study.

\subsection{The magnetic moments and form factors}
\label{subsec:mmom+mff}
The $^{3}$H and $^{3}$He magnetic form factors obtained by including 
nucleon and $\Delta$-isobar degrees of freedom in the nuclear wave 
functions and currents are shown in Figs.~\ref{fig:mff3h} and
~\ref{fig:mff3he}; while individual contributions to the combinations 
$F^{S}_{M}$ and $F^{V}_{M}$ are displayed in Figs.~\ref{fig:mffis_c} 
and~\ref{fig:mffiv_c}. Finally, individual and cumulative contributions 
to the magnetic moments and cumulative contributions to the 
magnetic radii of the trinucleons are listed in Tables
~\ref{tb:mmomcontr},~\ref{tb:mmom} and~\ref{tb:mgr}, 
respectively. Note that in Figs.~\ref{fig:mff3h} 
and~\ref{fig:mff3he} and Table~\ref{tb:mmomcontr} the 
contributions labelled 1-$\Delta$ and 2-$\Delta$ are associated with the 
diagrams in Figs.~\ref{fig:1bD} and~\ref{fig:2bD}, respectively. Also 
note that the individual nucleonic and $\Delta$-isobar contributions 
in Figs.~\ref{fig:mffis_c} and~\ref{fig:mffiv_c} and Table~\ref{tb:mmomcontr} 
are normalized as, in a schematic notation,
\begin{equation}
\left[ O \right] = \frac{\langle\Psi\,|\,j_{O}\,|\,\Psi\rangle}{\langle 
\Psi\,|\Psi\rangle} \ .
\label{eq:norm_c}
\end{equation}
However, the cumulative contributions in Figs.~\ref{fig:mff3h} and
~\ref{fig:mff3he} and Table~\ref{tb:mmom} and~\ref{tb:mgr} 
are normalized as
\begin{equation}
\left[ {\rm TOT\!-\!N} \right] = \frac{\langle\Psi\,|\,j(N\,{\rm only})\,|\,
\Psi\rangle}{\langle \Psi\,|\Psi\rangle} \ ,
\label{eq:norm_totN}
\end{equation}
when ``nucleons only'' terms are retained, and as
\begin{equation}
\left[ {\rm TOT\!-\!(N+\Delta)} \right] = \frac{\langle\Psi_{N+\Delta}\,|\,
j(N+\Delta)\,|\,\Psi_{N+\Delta}\rangle}{\langle \Psi_{N+\Delta}\,|
\Psi_{N+\Delta}\rangle} \ ,
\label{eq:norm_totND}
\end{equation}
when, in addition, the $\Delta$ terms are included.

The contributions associated with $\Delta$-components are found to be 
small in contrast to earlier studies~\cite{STR87}. In particular, we 
find that the sign of the 2-$\Delta$ contribution in Fig.~\ref{fig:mffiv_c} 
is opposite to that reported in Ref.~\cite{STR87}. The origin of this 
difference is unclear at this point. However, we do find that the sign 
of the 2-$\Delta$ contribution (see Fig.~\ref{fig:mffiv_c}) is the same 
as that of the nucleonic PS ($\pi$-like) contribution, as one would expect.

The predicted magnetic moments of the trinucleons are within less than 
1\% of the experimental values. The predominatly isovector $\Delta$-isobar  
contributions lead to an increase (in magnitude) of the $^{3}$H and 
$^{3}$He magnetic moments calculated with nucleons only degrees of 
freedom of, respectively, 1.1\% and 1.7\% in relative terms. We note 
that perturbation theory estimates of the $\Delta$-isobar 
contributions are found 
to be significantly larger than obtained here~\cite{SCH89}.

The predicted magnetic radii of $^{3}$H and $^{3}$He are, respectively, 
2\% and 3\% smaller than the experimental values, but still within 
experimental errors. Inclusion of the contributions due to 
two- and three-body exchange currents leads to a decrease of the 
$^{3}$H and $^{3}$He magnetic radii of, respectively, 
5\% and 6\%.

While the agreement between theory and experiment is satisfactory for the 
magnetic moments, magnetic radii and low $q$ form factors, the calculated 
form factors, particularly that of $^{3}$He, remain at variance with 
the experiment in the diffraction region. The role played by $\Delta$-isobar 
degrees of freedom is found to be marginal over the whole $q$-range 
considered here.

\section{Conclusions}
\label{sec:concl}

The present results for the electromagnetic form factors of the
trinucleons may be summarized as follows: i) the trinucleon charge form
factors agree well with the experimental values when calculated with 
wave functions obtained from a Hamiltonian consisting of the 
Argonne $v_{18}$ two-nucleon and the 
Urbana IX three-nucleon interactions; ii) agreement
with the experimental charge form factors 
requires that the two-nucleon
exchange charge operators are taken into account; iii) the calculated 
magnetic form factor of $^3$H agrees well with experiment, whereas
that of $^3$He agrees well with the experimental values only for
momentum transfer values below the first zero in the form factor; iv)
the two-nucleon exchange current contributions are essential for
achieving agreement with experiment whereas three-nucleon
exchange current operators and the $\Delta$-isobar configurations
have only very small effects on the calculated magnetic form factors.

The result for the static observables are that the calculated
value for the isovector combination of the trinucleon
magnetic moments agrees completely with the experimental
value (Table~\ref{tb:mmom}). The isoscalar combination of the trinucleon
magnetic moments exceeds the experimental value by about
5\%, but this small disagreement does not prevent a
good reproduction of the isoscalar combination of the experimental
magnetic form factors. As the calculated magnetic moments of
$^3$H and $^3$He differ by less than 0.015 n.m. from their
experimental values, the results appear to be very satisfactory.
The calculated charge radii are smaller by only 2\% than the
experimental values. The calculated magnetic radii are
smaller than 3\% than the experimental values 
(Tables~\ref{tb:mgr} and~\ref{tb:chr}).
To obtain these quite satisfactory calculated values for the charge
and magnetic radii the exchange current
contributions have to be taken into account.

We note finally that
the three-nucleon exchange current operator~(\ref{eq:WT9})--(\ref{eq:WT11}), 
which was constructed to satisfy the continuity equation
with the three-nucleon interaction~(\ref{eq:WT4}), was found
to give only very small contributions to the magnetic
form factors of the trinucleons. It is worth noting that the
corresponding component of the ``Tucson-Melbourne'' type
three-nucleon interaction in~\cite{COO93} is roughly an
order of magnitude stronger than the three-nucleon 
interaction~(\ref{eq:WT4}), and would therefore imply correspondingly
much larger three-nucleon exchange current contributions.
Part of this difference is the inclusion of $\Delta$-isobar
intermediate state effects in the ``d'' term of that three-nucleon
interaction. In the present work $\Delta$-isobar configurations
are treated explicitly, and should therefore not be
included in irreducible three-nucleon exchange current
operators. 

\acknowledgments
We wish to thank A. Kievsky, S. Rosati and M. Viviani for letting us use 
their PHH wave functions, and I. Sick for providing us with tables of the 
experimental data. The support of the U.S. Department of Energy via a 
graduate research assistantship provided by Jefferson Lab is gratefully 
acknowledged by L.E.M. The work of D.O.R. is partially
supported by the Academy of Finland under contract 34081 
while that of R.S. is supported by
the U.S. Department of Energy. Finally, the calculations 
were made possible by grants of time from the National Energy Research 
Supercomputer Center in Livermore.

\begin{table}
\caption{Binding energies corresponding to the AV14 and AV18/UIX 
Hamiltonian models. The AV14 results obtained with the PHH expansion 
are compared with those calculated by solving the Faddeev equations in 
configuration (F/R) and in momentum (F/P) space. The statistical 
error associated with the GFMC calculations are shown in 
parenthesis.}
\begin{tabular}{llll}
    Model  &   Method &   B($^{3}$H) (MeV)   &   B($^{3}$He) (MeV)   \\
\tableline 
           &    PHH   &        7.683          &    7.032         \\
     AV14  &  F/R     &        7.670          &    7.014         \\
           &  F/P     &        7.680          &                  \\
\tableline 
   AV18/UIX &    PHH   &        8.49           &    7.75          \\
            &    GFMC  &        8.47(1)        &    7.71(1)        \\
\tableline 
\multicolumn{2}{c}{expt.}
                      &        8.48        &    7.72          \\
\end{tabular}
\label{tb:binde}
\end{table}

\begin{table}
\caption{Individual contributions from the different components of the 
nuclear electromagnetic current operator to the $^{3}$H and $^{3}$He 
magnetic moments and their $\mu_{S}$ and $\mu_{V}$ combinations, in 
nuclear magnetons (n.m.). Note that, because of isospin-symmetry 
breaking components present in the PHH $^{3}$H and $^{3}$He wave functions, 
purely isoscalar (isovector) currents give non vanishing contributions 
to the $\mu_{V}$ ($\mu_{S}$) combination. 
The contributions to $\mu_{S}$ due to the $\pi\pi_S$ and 2-$\Delta$ 
currents and those to $\mu_{V}$ due to the SO2+LL currents are 
very small and are not listed.}
\begin{tabular}{lrrrr} 
            &  $\mu(^{3}$H) &  $\mu(^{3}$He) &  $\mu_S$    & $\mu_V$    \\
\tableline
1-N         &    2.571       &    --1.757     &     0.407     &  2.164    \\
PS          &    0.274       &    --0.269     &     0.002     &  0.271    \\
V           &    0.046       &    --0.044     &     0.001     &  0.045    \\
SO          &    0.057       &      0.010     &     0.033     &  0.023    \\
SO2+LL      &   --0.005      &    --0.006     &   --0.005     &           \\
$\rho\pi\gamma$+$\omega\pi\gamma$ &    
                 0.016      &    --0.009     &    0.003     &  0.012    \\
$\pi\pi_S$  &    0.002      &    --0.002     &              &  0.002    \\
1-$\Delta$  &    0.084      &    --0.064     &    0.010     &  0.074    \\
2-$\Delta$  &    0.024      &    --0.024     &              &  0.024    \\
\end{tabular}
\label{tb:mmomcontr}
\end{table}

\begin{table}
\caption{Cumulative and normalized contributions to the $^{3}$H and 
$^{3}$He magnetic moments and their $\mu_S$ and $\mu_V$ combinations, 
in nuclear magnetons (n.m.), compared with the experimental data.}
\begin{tabular}{lrrrr} 
            &  $\mu(^{3}$H) &  $\mu(^{3}$He) &  $\mu_S$    & $\mu_V$        \\
\tableline
1-N         &    2.571      &    --1.757     &    0.407     &  2.164    \\
TOT-N         
            &    2.961      &    --2.077     &    0.442     &  2.519    \\
TOT-N+1-$\Delta$  
            &    2.971      &    --2.089     &    0.441     &  2.530    \\
TOT-(N+$\Delta$)         
            &    2.994      &    --2.112     &    0.441     &  2.553    \\
EXP         &    2.979      &    --2.127     &    0.426     &  2.553    \\
\end{tabular}
\label{tb:mmom}
\end{table}

\begin{table}
\caption{Cumulative and normalized contributions to the $^{3}$H and 
$^{3}$He r.m.s. magnetic radii, in fm, compared with the experimental 
data.}
\begin{tabular}{lcc} 
            &  $^{3}$H &  $^{3}$He   \\
\tableline
1-N         &    1.895                &    2.040             \\
TOT-N
            &    1.810                &    1.925             \\
TOT-N+1-$\Delta$  
            &    1.804                &    1.916             \\
TOT-(N+$\Delta$)         
            &    1.800                &    1.909             \\
EXP         &    1.840$\pm$0.181      &    1.965$\pm$0.153   \\
\end{tabular}
\label{tb:mgr}
\end{table}

\begin{table}
\caption{Cumulative and normalized contributions to the $^{3}$H and 
$^{3}$He r.m.s. charge radii, in fm, compared with the 
experimental data.}
\begin{tabular}{lcc} 
            &  $^{3}$H                &  $^{3}$He   \\
\tableline
1-N         &    1.711                &    1.919             \\
TOT         &    1.725                &    1.928             \\
EXP         &    1.755$\pm$0.086      &    1.959$\pm$0.030   \\
\end{tabular}
\label{tb:chr}
\end{table}
\begin{figure}[!hbt]
\caption[Figure]{\label{fig:mff3hNN} The magnetic form factors of $^{3}$H, 
obtained with single-nucleon currents (1-N), and with inclusion of 
two-nucleon current [(1+2)-N] and $\pi\pi_S$ three-nucleon 
[TOT-N(D)] current contributions, are compared with data (shaded area) 
from Amroun $et$ $al.$~\cite{AMR94}. Theoretical results correspond to 
the AV18/UIX PHH wave functions, and employ the dipole parametrization 
(including the Galster factor for $G_{E}(q_{\mu}^{2})$) for the nucleon 
electromagnetic form factors. Note that the Sachs form factor 
$G_{E}^{V}(q_{\mu}^{2})$ is used in the model-independent isovector 
two-body currents obtained from the charge-independent part of the AV18 
interaction. Also shown are the total results corresponding 
to the Gari-Kr\"{u}mpelmann parametrization~\cite{GAR86} of the nucleon 
electromagnetic form factor [TOT-N(GK)].}
\end{figure}
\begin{figure}[!hbt]
\caption[Figure]{\label{fig:mff3heNN} Same as in Fig.~\ref{fig:mff3hNN}, 
but for $^{3}$He.}
\end{figure}
\begin{figure}[!hbt]
\caption[Figure]{\label{fig:mffis_cNN} Individual contributions to the 
$F_{M}^{S}(q_{\mu})$ combination, Eq.~(\ref{eq:isvscmff}), of the $^{3}$H 
and $^{3}$He magnetic form factors, obtained with the dipole parametrization 
of the nucleon electromagnetic form factors. The sign of each contribution 
is given in parenthesis. Note that, because of isospin-symmetry breaking 
components present in the $^{3}$H and $^{3}$He wave functions, the purely 
isovector PS, V and $\pi\pi_S$ currents give non vanishing 
contributions to the $F_{M}^{S}(q_{\mu})$ combination. However as the
$\pi\pi_S$
contribution is very small, is not shown.}
\end{figure}
\begin{figure}[!hbt]
\caption[Figure]{\label{fig:mffiv_cNN}Individual contributions to the 
$F_{M}^{V}(q_{\mu})$ combination, Eq.~(\ref{eq:isvscmff}), of the $^{3}$H 
and $^{3}$He magnetic form factors, obtained with the dipole parametrization 
of the nucleon electromagnetic form factors. The sign of each contribution 
is given in parenthesis. Note that, because of isospin-symmetry breaking 
components present in the $^{3}$H and $^{3}$He wave functions, the purely 
isoscalar $\rho\pi\gamma$ current gives non vanishing 
contributions to the $F_{M}^{V}(q_{\mu})$ combination. However, 
being very small, it is not shown.}
\end{figure}
\begin{figure}[!hbt]
\caption[Figure]{\label{fig:mffisNN} The $F_{M}^{S}(q_{\mu})$ combinations 
of the $^{3}$H and $^{3}$He magnetic form factors, 
obtained with single-nucleon 
currents (1-N), and with inclusion of two-nucleon current [(1+2)-N] and 
$\pi\pi_S$ three-nucleon current (TOT-N) contributions, are compared 
with data (shaded area) from Amroun $et$ $al.$~\cite{AMR94}. The dipole 
parametrization is used for the nucleon electromagnetic form factors.}
\end{figure}
\begin{figure}[!hbt]
\caption[Figure]{\label{fig:mffivNN} Same as in Fig.~\ref{fig:mffisNN}, 
but for the $F_{M}^{V}(q_{\mu})$ combination of the $^{3}$H and $^{3}$He 
magnetic form factors.}
\end{figure}
\begin{figure}[!hbt]
\caption[Figure]{\label{fig:cff3hNN} The charge form factors of $^{3}$H, 
obtained with a single-nucleon charge operator (1-N) and with inclusion 
of two-nucleon charge operator contributions (TOT-N), are compared with data 
(shaded area) from Amroun $et$ $al.$~\cite{AMR94}. Note that the 1-N results 
also include the Darwin-Foldy and spin-orbit corrections. Theoretical results 
correspond to the AV18/UIX PHH wave functions, and employ the dipole 
parametrization of the nucleon electromagnetic form factors.}
\end{figure}
\begin{figure}[!hbt]
\caption[Figure]{\label{fig:cff3heNN} Same as in Fig.~\ref{fig:cff3heNN}, 
but for $^{3}$He.}
\end{figure}
\begin{figure}[!hbt]
\caption[Figure]{\label{fig:cffis_cNN} Individual contributions to the 
$F_{C}^{S}(q_{\mu})$ combination, Eq.~(\ref{eq:isvsccff}), of the $^{3}$H 
and $^{3}$He charge form factors, obtained with the dipole parametrization 
of the nucleon electromagnetic form factors. The sign of each combination 
is given in parenthesis. Note that, because of isospin-symmetry breaking 
components present in the $^{3}$H and $^{3}$He wave functions, the purely 
isovector $\omega\pi\gamma$ charge operator gives a non vanishing 
contribution to the $F_{C}^{S}(q_{\mu})$ combination.}
\end{figure}
\begin{figure}[!hbt]
\caption[Figure]{\label{fig:cffiv_cNN} Individual contributions to the 
$F_{C}^{V}(q_{\mu})$ combination, Eq.~(\ref{eq:isvsccff}), of the $^{3}$H 
and $^{3}$He charge form factors, obtained with the dipole parametrization 
of the nucleon electromagnetic form factors. The sign of each combination 
is given in parenthesis. Note that, because of isospin-symmetry breaking 
components present in the $^{3}$H and $^{3}$He wave functions, the purely 
isoscalar $\rho\pi\gamma$ charge operator gives a non vanishing 
contribution to the $F_{C}^{V}(q_{\mu})$ combination.}
\end{figure}
\begin{figure}[!hbt]
\caption[Figure]{\label{fig:tcf}Transition correlation functions 
         $u^{\sigma\tau II}(r)$, $u^{t\tau II}(r)$, etc. 
         obtained for the AV28Q model~\cite{WIRpc}, 
         and perturbation theory 
         equivalents $u^{\sigma\tau II, {\rm PT}}(r)$,
         $u^{t\tau II, {\rm PT}}(r)$, etc.}
\end{figure}
%
%
%\newpage
\begin{figure}[!hbt]
\caption[Figure]{\label{fig:cur2Delta}$N\Delta$-transition 
two-body currents due to pion exchange.}
\end{figure}
%\newpage
%
%
\begin{figure}[!hbt]
\caption[Figure]{\label{fig:1bD}Diagrammatic representation of operators 
included in $j(\Delta)$ due to one-body currents $j^{(1)}
(N\rightarrow\Delta)$, $j^{(1)}(\Delta\rightarrow N)$ and $j^{(1)}
(\Delta\rightarrow\Delta)$, and transition correlations $U^{N\Delta}$, 
$U^{\Delta N}$, $U^{\Delta\Delta}$, and corresponding hermitian conjugates. 
Wavy, thin, thick, dashed and cross-dashed lines denote photons, nucleons, 
$\Delta$-isobars and transition correlations $U^{B B^{'}}$ and 
${U^{B B^{'}}}^{\dagger}$, respectively.}
\end{figure}
\begin{figure}[!hbt]
\caption[Figure]{\label{fig:2bD}Diagrammatic representation of operators 
included in $j(\Delta)$ due to two-body currents $j^{(2)}
(NN\rightarrow N\Delta)$, $j^{(2)}(NN\rightarrow \Delta N)$, etc., and 
transition correlations $U^{N\Delta}$, $U^{\Delta N}$, 
and corresponding hermitian conjugates. Wavy, thin, thick, dashed and 
cross-dashed lines denote photons, nucleons, $\Delta$-isobars and 
transition correlations $U^{B B^{'}}$ and ${U^{B B^{'}}}^{\dagger}$, 
respectively.}
\end{figure}
%
%
%\newpage
\begin{figure}[!hbt]
\caption[Figure]{\label{fig:3bDnegl} Diagrams associated with connected 
three-body terms, which are neglected in the present work. Wavy, thin, 
thick, dashed, cross-dashed and dotted lines denote photons, nucleons, 
$\Delta$-isobars, transition correlations $U^{B B^{'}}$ and 
${U^{B B^{'}}}^{\dagger}$, and the two-body current 
$j^{(2)}(NN\rightarrow NN)$, respectively.}
\end{figure}
%\newpage
%
%
\begin{figure}[!hbt]
\caption[Figure]{\label{fig:mff3h} The magnetic form factors of $^{3}$H, 
obtained with single-nucleon currents (1-N), and with inclusion of 
two- and three-nucleon current (TOT-N) and $\Delta$ [TOT-(N+$\Delta$)] 
contributions.}
\end{figure}
\begin{figure}[!hbt]
\caption[Figure]{\label{fig:mff3he} Same as in Fig.~\ref{fig:mff3h}, 
but for $^{3}$He.}
\end{figure}
\begin{figure}[!hbt]
\caption[Figure]{\label{fig:mffis_c} The single-nucleon contribution to 
the $F_{M}^{S}(q_{\mu})$ combination of the $^{3}$H and $^{3}$He magnetic 
form factors is compared with the 1-$\Delta$ and 2-$\Delta$ contributions, 
associated respectively with diagrams of Fig.~\ref{fig:1bD} 
and~\ref{fig:2bD}.}
\end{figure}
\begin{figure}[!hbt]
\caption[Figure]{\label{fig:mffiv_c} The single-nucleon and leading PS 
two-nucleon contributions to the $F_{M}^{V}(q_{\mu})$ combination 
of the $^{3}$H and $^{3}$He magnetic form factors are compared with the 
1-$\Delta$ and 2-$\Delta$ contributions, associated respectively with 
diagrams of Fig.~\ref{fig:1bD} and~\ref{fig:2bD}.}
\end{figure}

\begin{references}
\bibitem{CAR98} J.\ Carlson and R.\ Schiavilla, preprint JLAB-THY-97-20, 
Rev.\ Mod.\ Phys.\ in press (1998).
\bibitem{PUD97} B.S.\ Pudliner, V.R.\ Pandharipande, J.\ Carlson,
S.C.\ Pieper, and R.B.\ Wiringa, Phys.\ Rev.\ C {\bf 56}, 1720 (1997).
\bibitem{WIR95} R.B.\ Wiringa, V.G.J.\ Stoks, and R.\ Schiavilla, Phys.\ 
Rev.\ C {\bf 51}, 38 (1995).
\bibitem{PUD95} B.S.\ Pudliner, V.R.\ Pandharipande, J.\ Carlson, 
and R.B.\ Wiringa, Phys.\ Rev.\ Lett. {\bf 74}, 4396 (1995).
\bibitem{SCH89} R.\ Schiavilla, V.R.\ Pandharipande, and D.O.\ Riska, 
Phys.\ Rev.\ C {\bf 40}, 2294 (1989).
\bibitem{SCH90} R.\ Schiavilla, V.R.\ Pandharipande, and D.O.\ Riska,
Phys.\ Rev.\ C {\bf 41}, 309 (1990).
\bibitem{WIR84} R.B.\ Wiringa, R.A.\ Smith, and T.L.\ Ainsworth, Phys.\ 
Rev.\ C {\bf 29}, 1207 (1984).
\bibitem{PEN90} M.T.\ Pe\~{n}a, H.\ Henning and P.U.\ Sauer,
Phys.\ Rev.\ C {\bf 42}, 855 (1990).
\bibitem{WIR88} R.B.\ Wiringa, V.\ Ficks, and A.\ Fabrocini, Phys.\ 
Rev.\ C {\bf 38}, 1010 (1988).
\bibitem{KIE93} A.\ Kievsky, S.\ Rosati, and M.\ Viviani, Nucl.\ Phys. 
{\bf A551}, 241 (1993).
\bibitem{KIE94} A.\ Kievsky, M.\ Viviani, and S.\ Rosati, Nucl.\ Phys. 
{\bf A577}, 511 (1994).
\bibitem{KIE95} A.\ Kievsky, M.\ Viviani, and S.\ Rosati, Phys.\ Rev.\ C 
{\bf 52}, R15 (1995).
\bibitem{VIV95} M.\ Viviani, A.\ Kievsky, and S.\ Rosati, Few-Body\ Syst. 
{\bf 18}, 25 (1995).
\bibitem{GLO83} W.\ Gl\"{o}ckle, {\it The Quantum Mechanical Few-Body 
Problem} (Springer-Verlag, Berlin-Heiderberg, 1983).
\bibitem{CHE86} C.R.\ Chen, C.L.\ Payne, J.L.\ Friar, and B.F.\ Gibson,
Phys.\ Rev.\ C {\bf 33}, 1740 (1986).
\bibitem{GLO95} W.\ Gl\"{o}ckle $et\, al.$, Few\ Body\ Syst.\ Suppl. 
{\bf 8}, 9 (1995).
\bibitem{RIS89} D.O.\ Riska, Phys.\ Rep. {\bf 181}, 207 (1989).
\bibitem{VIV96} M.\ Viviani, R.\ Schiavilla, and A.\ Kievsky, Phys.\ Rev.\ 
C {\bf 54}, 534 (1996).
\bibitem{BER80} H.\ Berg $et\, al.$, Nucl.\ Phys. {\bf A334}, 21 (1980).
\bibitem{CHE71-73} M.\ Chemtob and M.\ Rho, Nucl.\ Phys. {\bf A163}, 
1 (1971); {\bf A212}, 628 (1973).
\bibitem{CAR91} J.\ Carlson, V.R.\ Pandharipande, and R.\ Schiavilla, 
{\it Modern Topics in Electron Scattering}, B.\ Frois and I.\ Sick, 
eds. (World Scientific, Singapore, 1991), p. 177.
\bibitem{RIS85} D.O.\ Riska,\ Phys.\ Scr. {\bf 31}, 471 (1985).
\bibitem{CAR90} J.\ Carlson, D.O.\ Riska, R.\ Schiavilla, and 
R.B.\ Wiringa, Phys.\ Rev.\ C {\bf 42}, 830 (1990).
\bibitem{SCH91} R.\ Schiavilla and D.O.\ Riska, 
Phys.\ Rev.\ C {\bf 43}, 437 (1991). 
\bibitem{FRI77} J.L.\ Friar, Ann.\ Phys. {\bf 104}, 380 (1977).
\bibitem{COO86} S.A.\ Coon and J.L.\ Friar, 
Phys.\ Rev.\ C {\bf 34}, 1060 (1986).
\bibitem{COE94} F.\ Coester and D.O.\ Riska, Ann.\ Phys. {\bf 234}, 141
(1994).
\bibitem{SCHtbp} R.\ Schiavilla, to be published.
\bibitem{WEI67} S.\ Weinberg, Phys.\ Rev.\ Lett. {\bf 18}, 188 (1967).
\bibitem{COO93} S.A.\ Coon and M.T.\ Pe\~{n}a, Phys.\ Rev.\ C
{\bf 48}, 2559 (1993).
\bibitem{MET53} N.\ Metropolis $et\, al.$, 
J.\ Chem.\ Phys. {\bf 21}, 1087 (1953).
\bibitem{COL65} H.\ Collard $et\, al.$, Phys.\ Rev.\ {\bf 138}, 357 (1965).
\bibitem{MCC77} J.S.\ McCarthy, I.\ Sick, and R.\ Whitney, 
Phys.\ Rev.\ C {\bf 15}, 1396 (1977).
\bibitem{SZA77} Z.M.\ Szalata $et\, al.$, Phys.\ Rev.\ C {\bf 15}, 1200 (1977).
\bibitem{ARN78} R.G.\ Arnold $et\, al.$, Phys.\ Rev.\ Lett. {\bf 40}, 
1429 (1978).
\bibitem{CAV82} J.M.\ Cavedon $et\, al.$, Phys.\ Rev.\ Lett. {\bf 49},
978 (1982).
\bibitem{DUN83} P.C.\ Dunn $et\, al.$, Phys.\ Rev.\ C {\bf 27}, 71 (1983).
\bibitem{OTT85} C.R.\ Ottermann $et\, al.$, Nucl.\ Phys.\ 
{\bf A435}, 688 (1985).
\bibitem{JUS85} F.P.\ Juster $et\, al.$, Phys.\ Rev.\ Lett. {\bf 55}, 
2261 (1985).
\bibitem{BEC87} D.H.\ Beck $et\, al.$, Phys.\ Rev.\ Lett. {\bf 59}, 
1537 (1987).
\bibitem{AMR94} A.\ Amroun $et\, al.$, Nucl.\ Phys.\ 
{\bf A579}, 596 (1994).
\bibitem{HOC73} J.\ Hockert, D.O.\ Riska, M.\ Gari, and A.\ Huffman,
Nucl.\ Phys.\ {\bf A217}, 14 (1973).
\bibitem{GAR86} M.\ Gari and W.\ Kr\"{u}mpelmann, Phys.\ Lett. 
{\bf B173}, 10 (1986).
\bibitem{SCH96} R.\ Schiavilla, {\it Perspectives in Nuclear Physics 
at Intermediate Energies}, S.\ Boffi, C.\ Ciofi degli Atti, and 
M.\ Giannini, eds., (World Scientific, Singapore, 1996).
\bibitem{HAJ83} Ch.\ Hajduk, P.U.\ Sauer, and W.\ Strueve, Nucl.\ Phys.\ 
{\bf A405}, 581 (1983).
\bibitem{PIC92} A.\ Picklesimer, R.A.\ Rice, and R.\ Brandenburg, Phys.\ 
Rev.\ Lett.\ {\bf 68}, 1484 (1992).
\bibitem{LOM81} J.\ Lomnitz-Adler, V.R.\ Pandharipande, and R.A.\ Smith, 
Nucl.\ Phys.\ {\bf A361}, 399 (1981).
\bibitem{WIR91} R.B.\ Wiringa, Phys.\ Rev.\ C {\bf 43}, 1585 (1991).
\bibitem{SCH92} R.\ Schiavilla, R.B.\ Wiringa, V.R.\ Pandharipande, and J.\ 
Carlson, Phys.\ Rev.\ C {\bf 45}, 2628 (1992).
\bibitem{WIRpc} R.B.\ Wiringa, private comunication.
\bibitem{CAR86} C.E.\ Carlson, Phys.\ Rev.\ D {\bf 34}, 2704 (1986).
\bibitem{LIN91} D.\ Lin and M.K.\ Liou, Phys.\ Rev.\ C {\bf 43}, R930 
(1991).
\bibitem{ERI88} T.E.O.\ Ericson and W.\ Weise, {\it Pions and Nuclei} 
(Clarendon Press, Oxford, 1988).
\bibitem{FUJ57} J.\ Fujita and H.\ Miyazawa, Prog.\ Theor.\ Phys.\ 
{\bf 17}, 360 (1957).
\bibitem{STR87} W.\ Strueve, Ch.\ Hajduk, P.U.\ Sauer, and W.\ Theis, 
Nucl.\ Phys.\ {\bf A465}, 651 (1986).
\end{references}
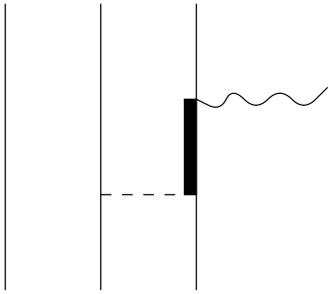
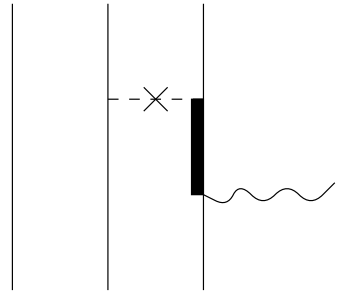
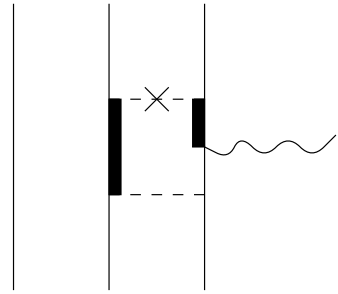
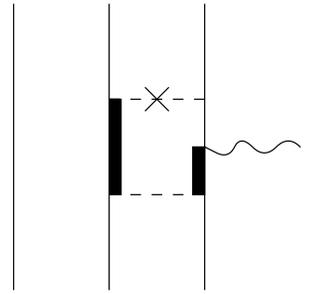
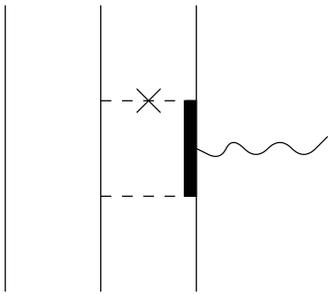
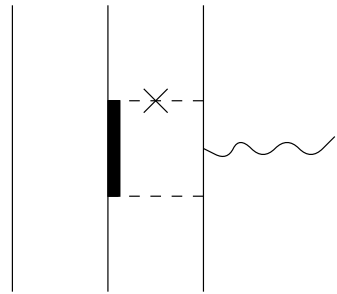
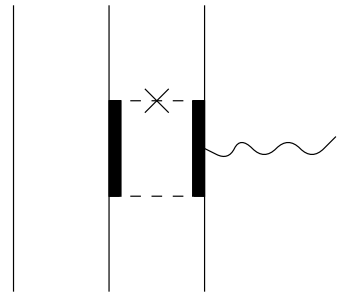
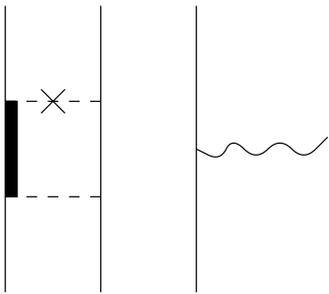
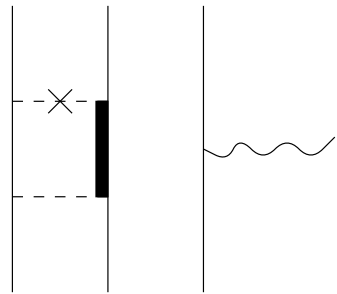
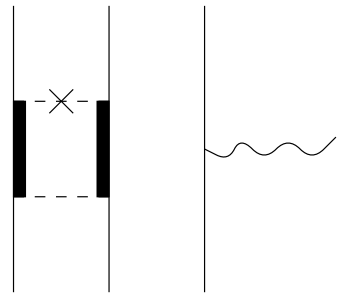
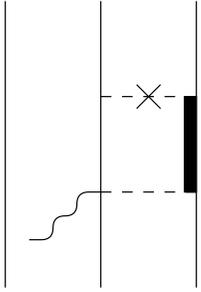
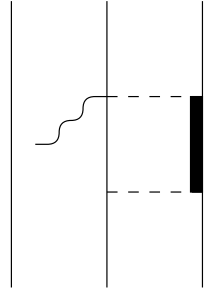
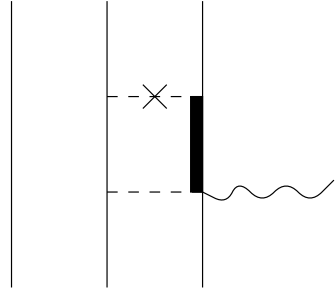
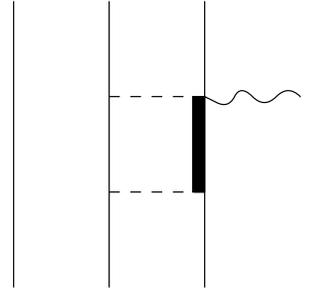
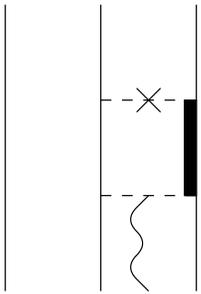
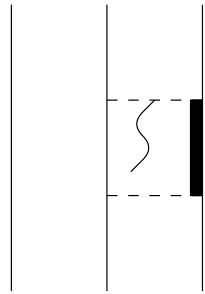
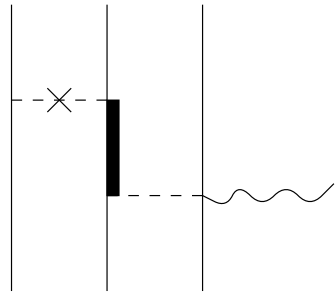
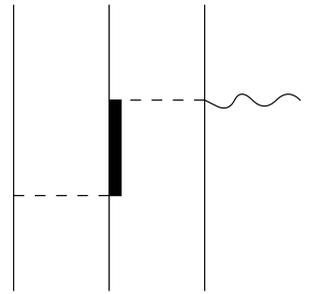
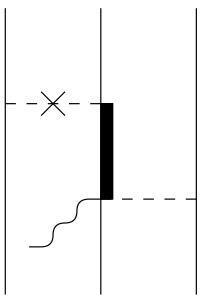
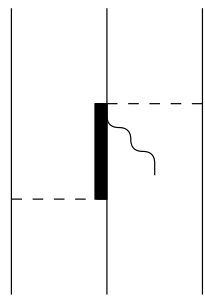
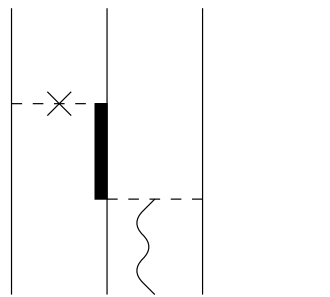
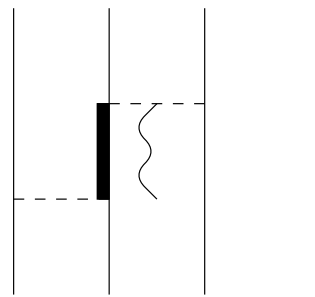
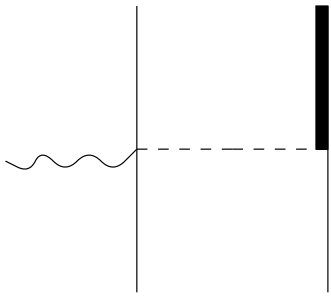
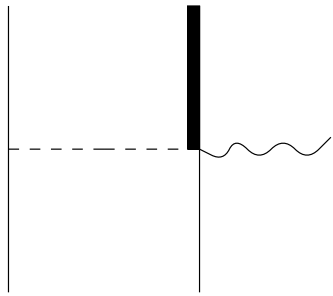
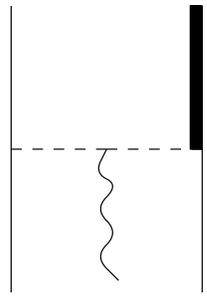
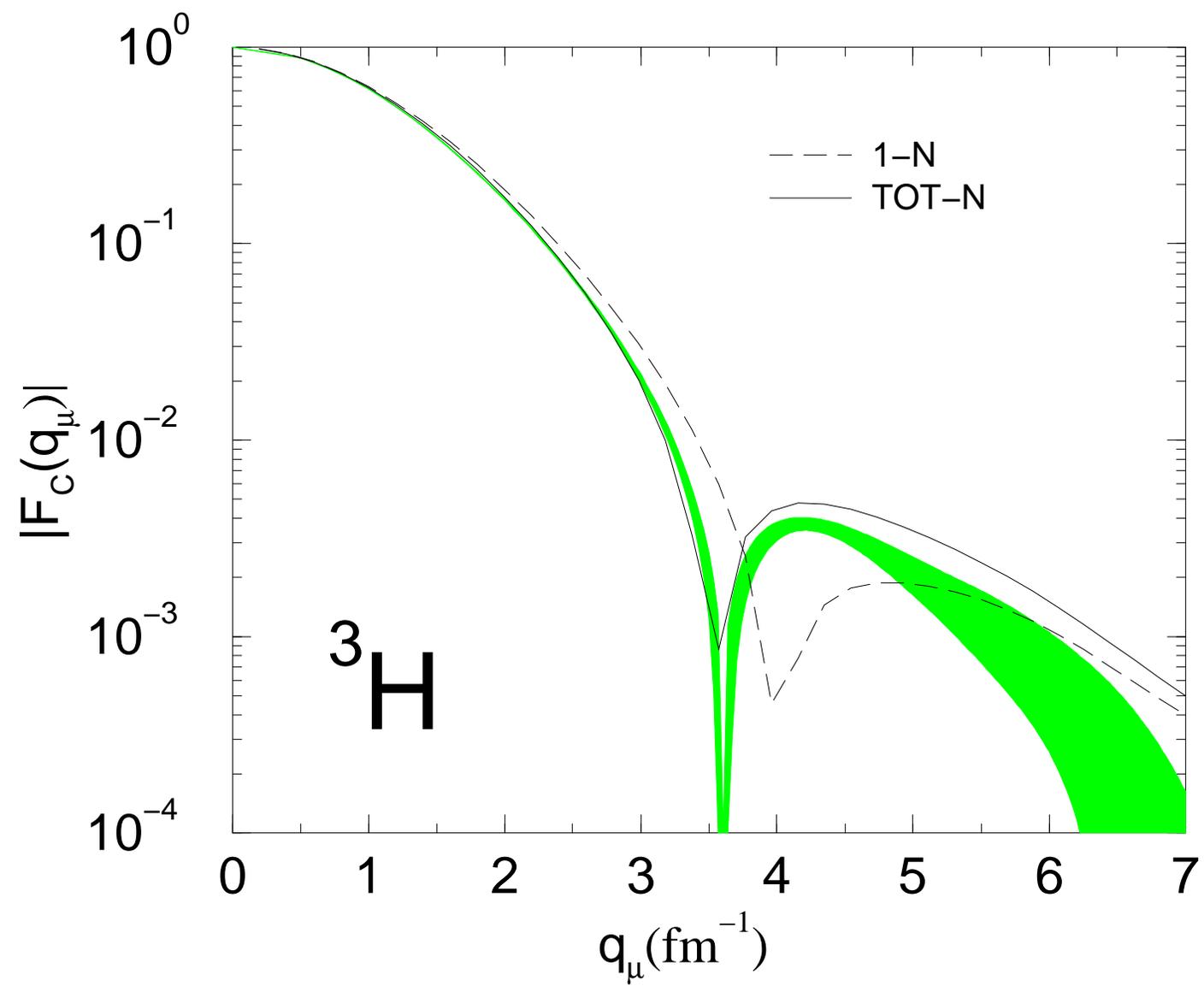
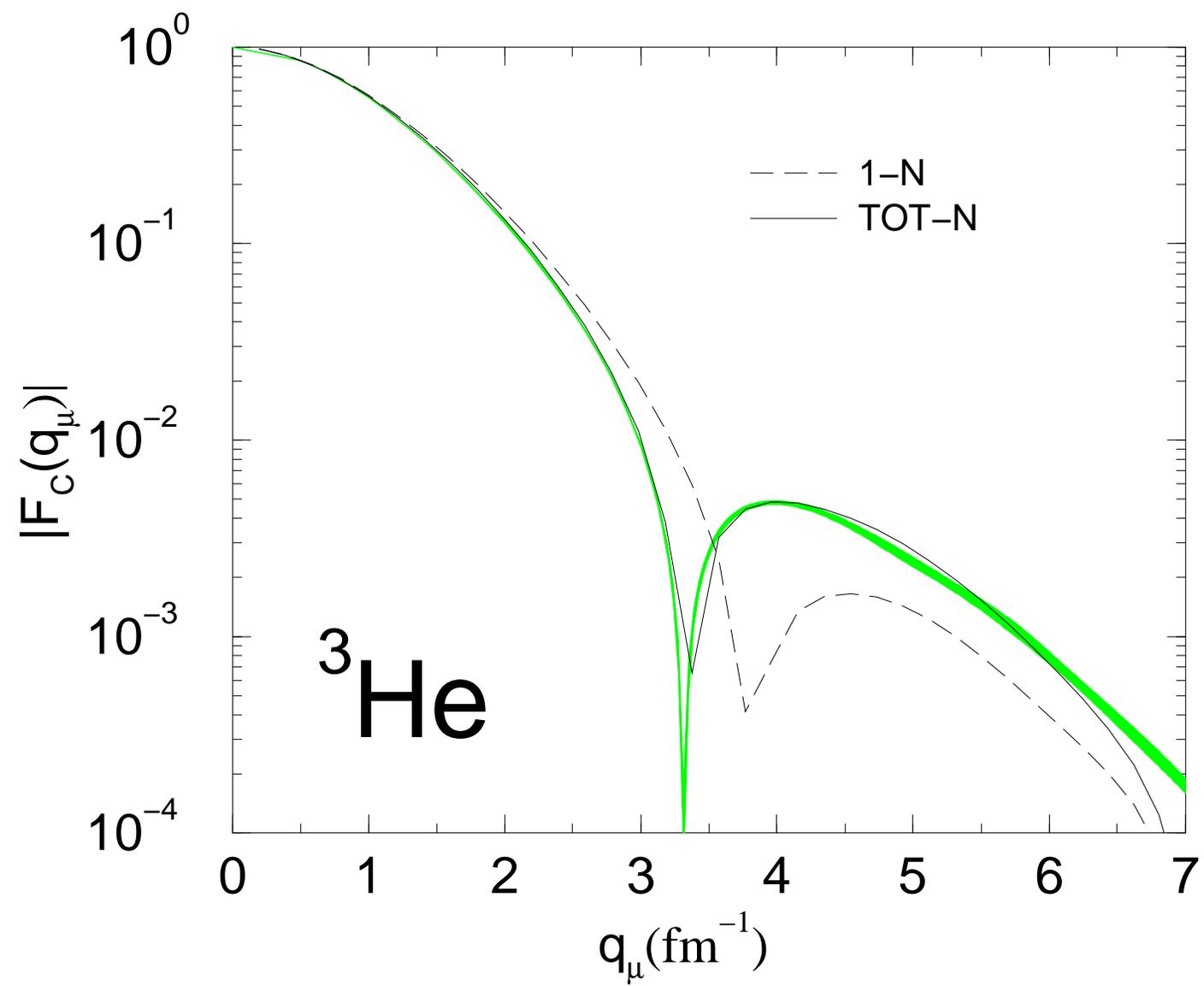
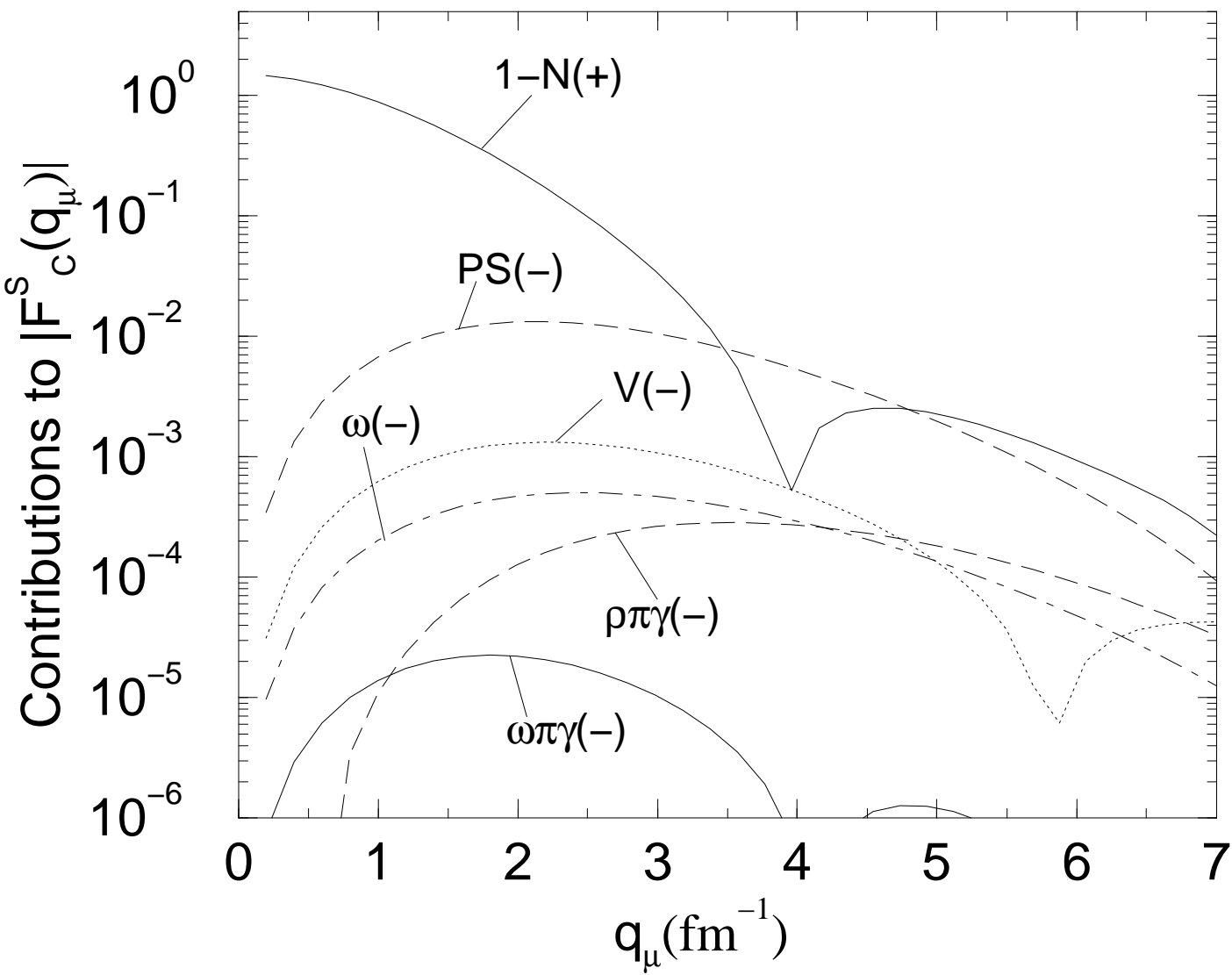
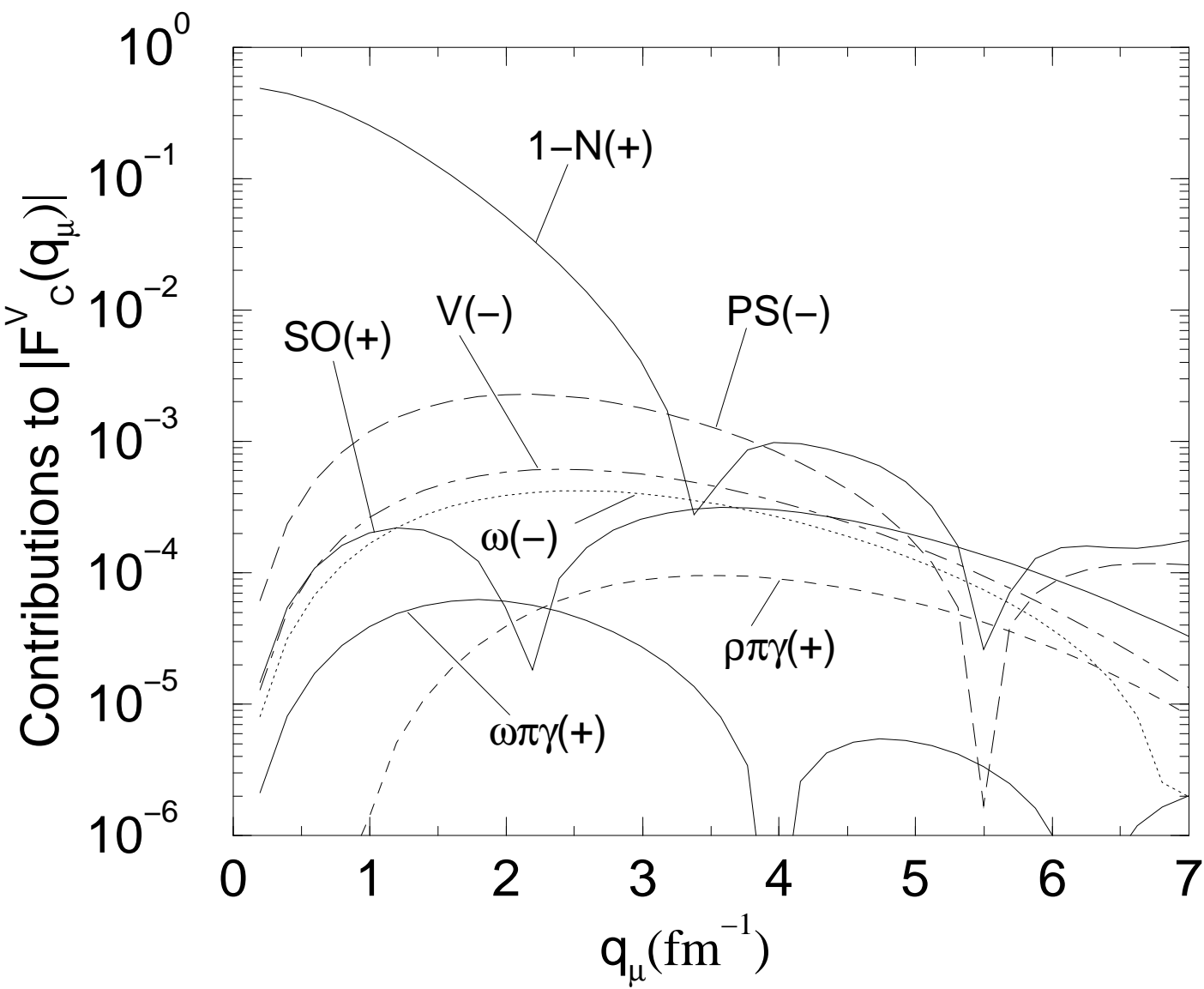
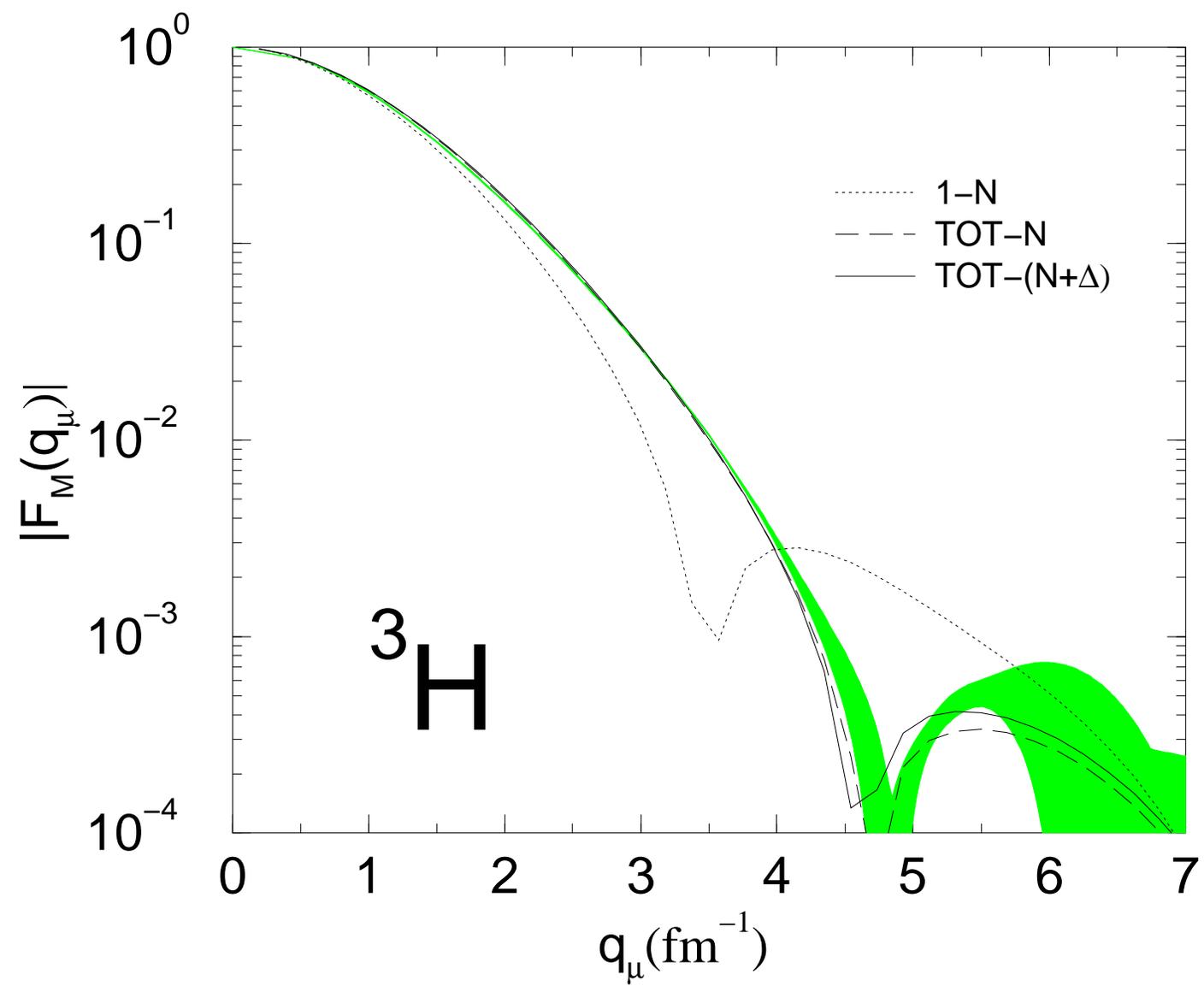
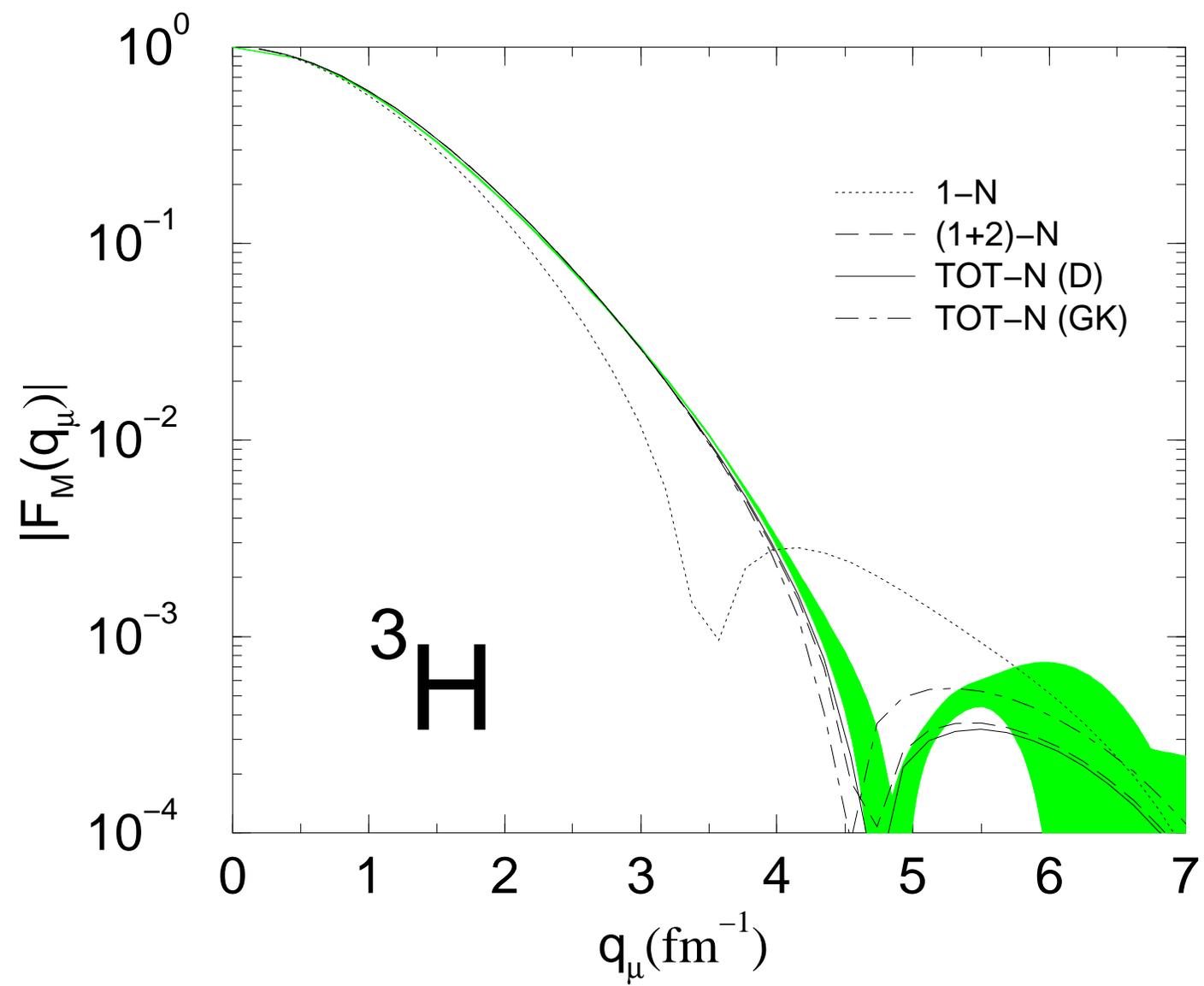
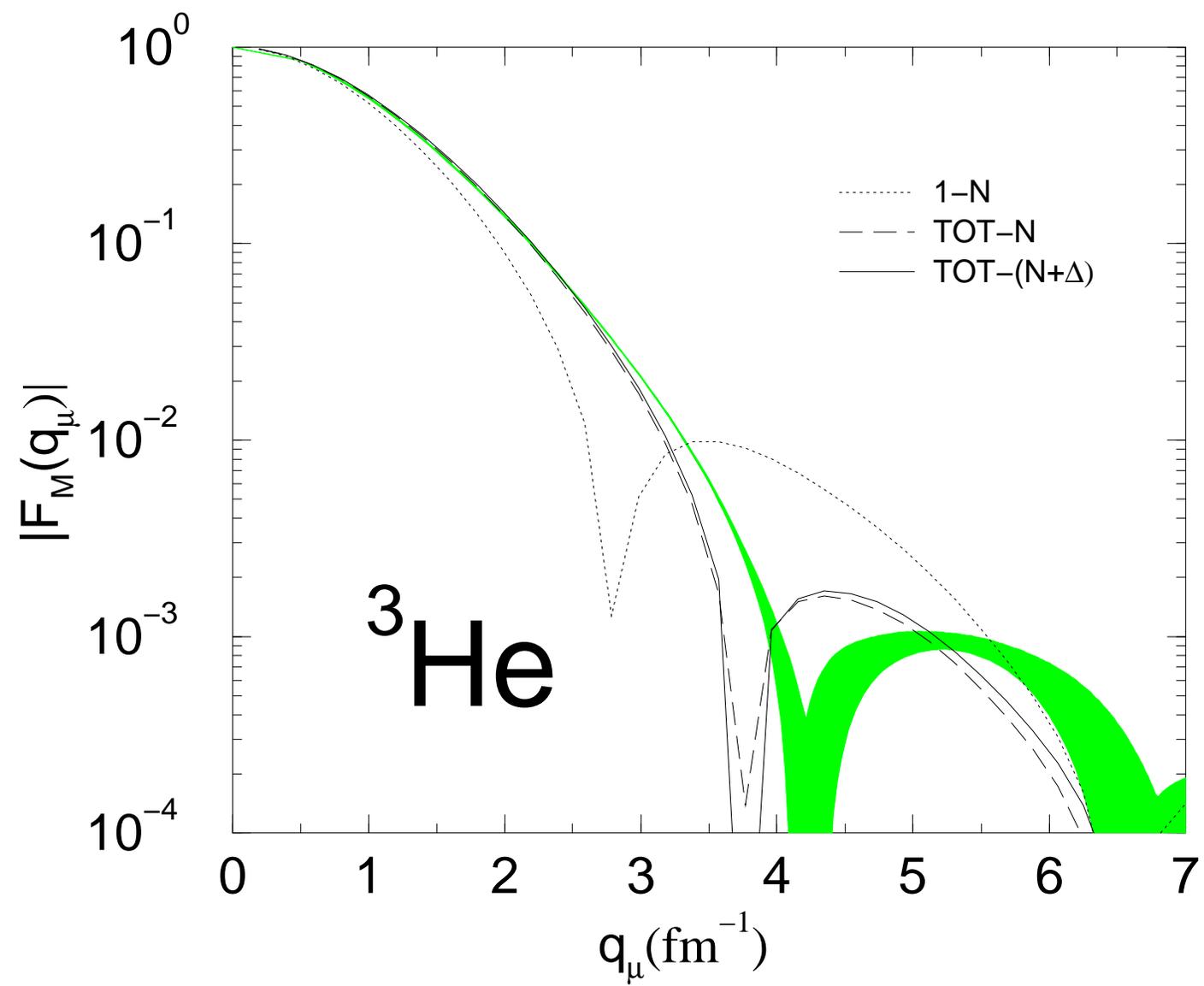
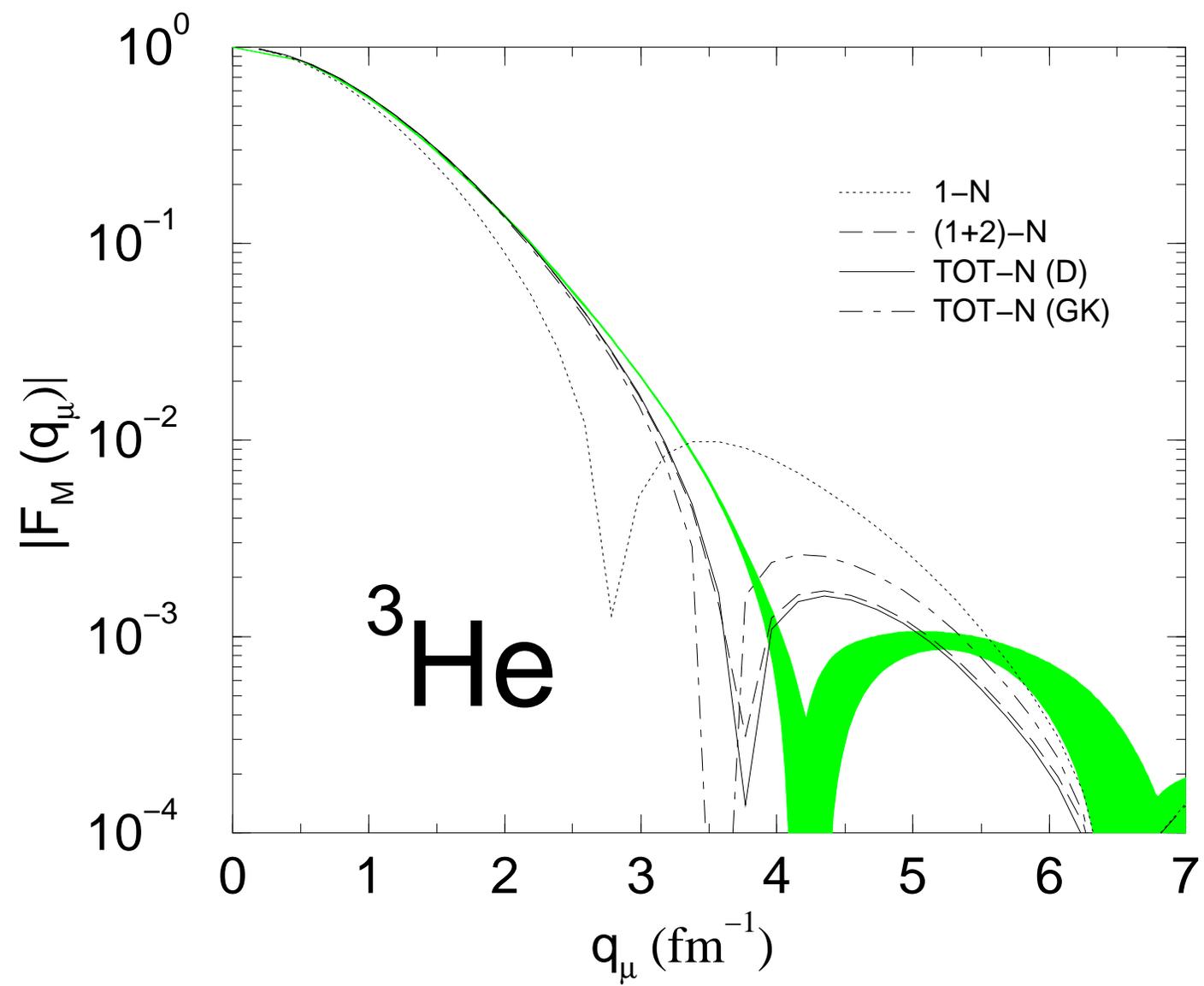
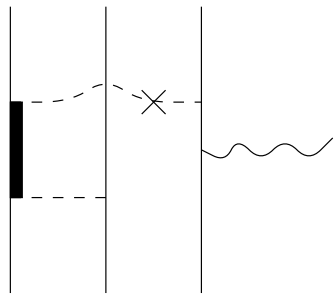
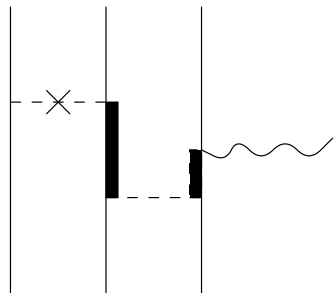
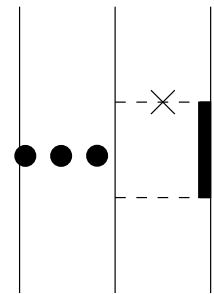
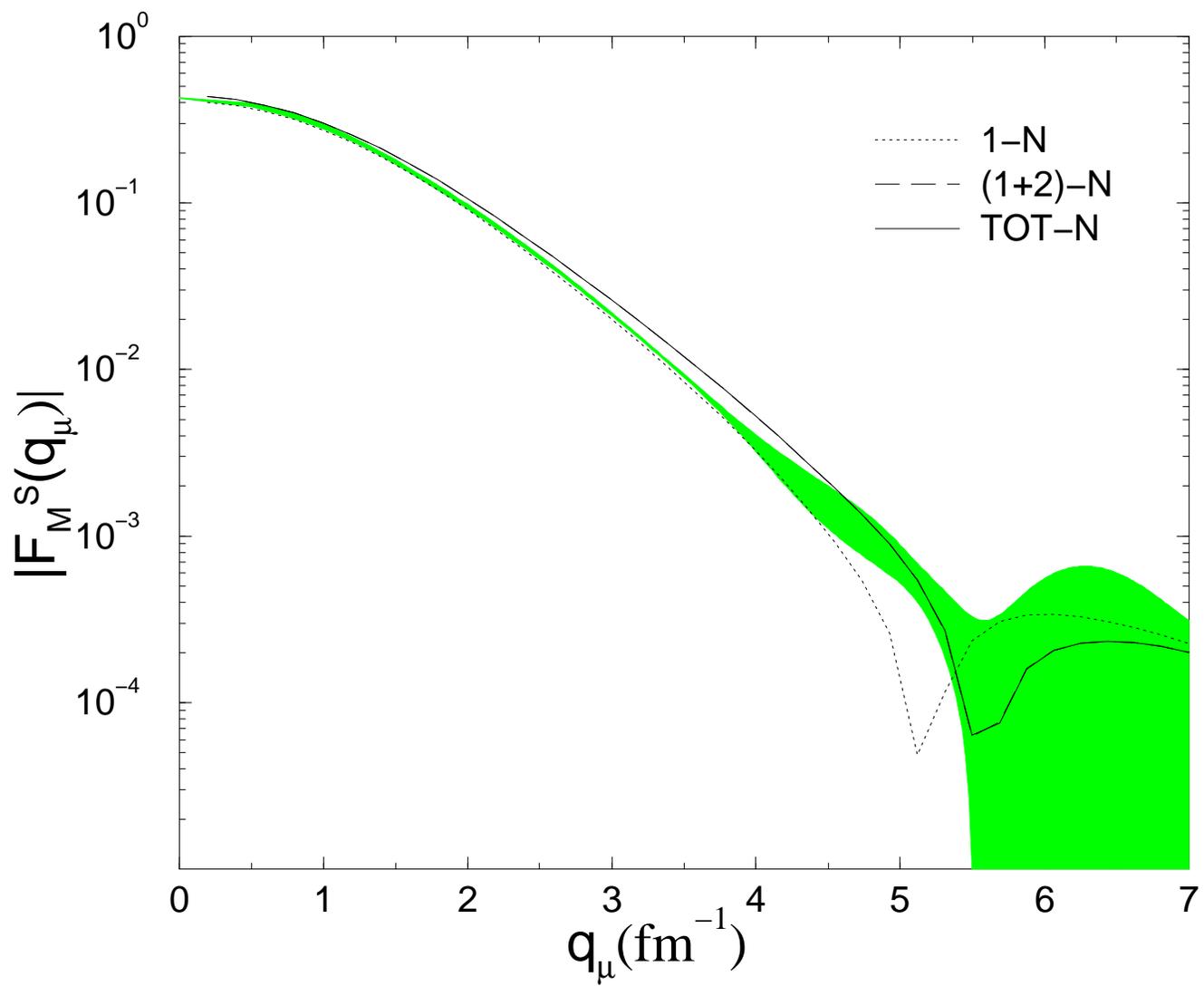
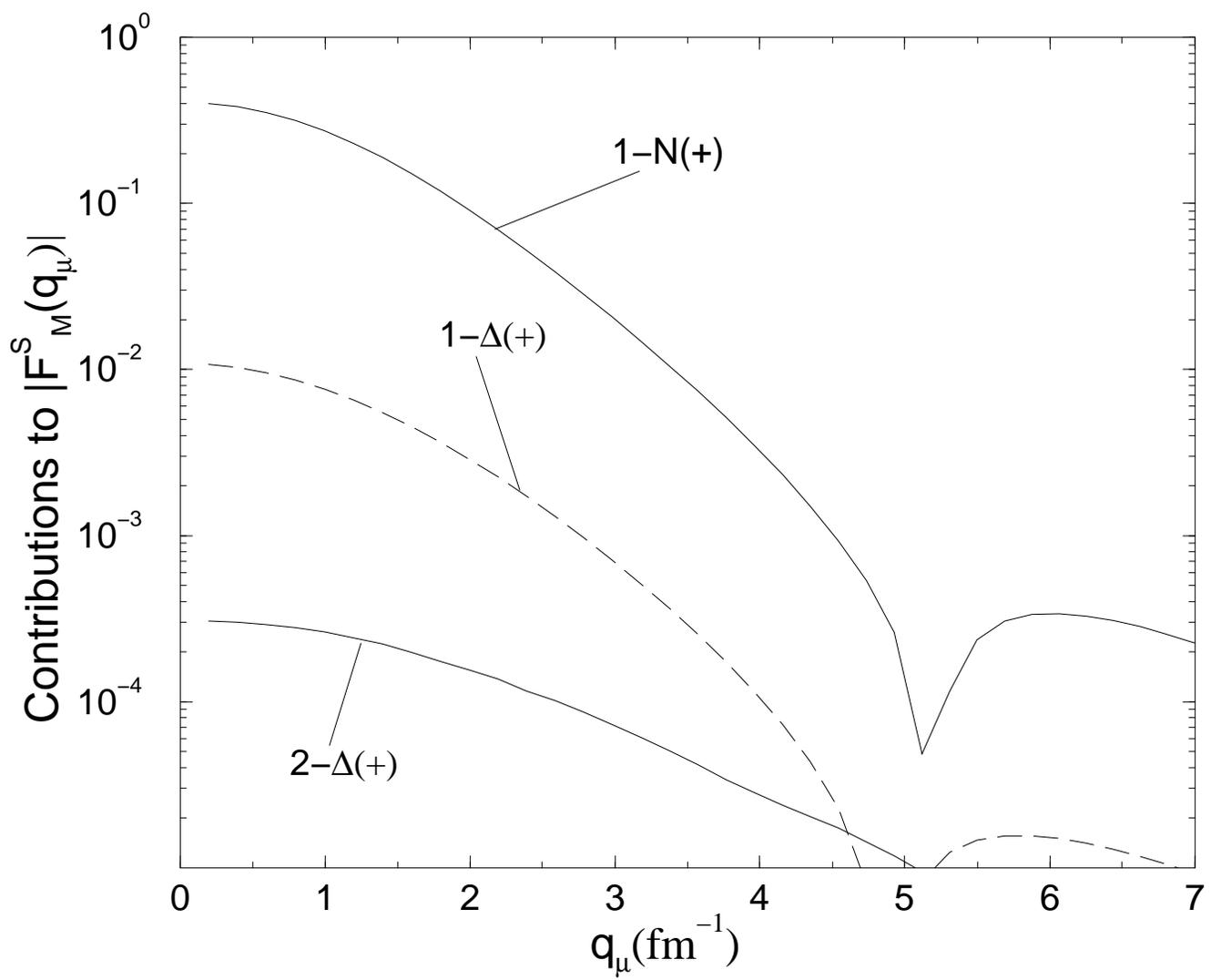
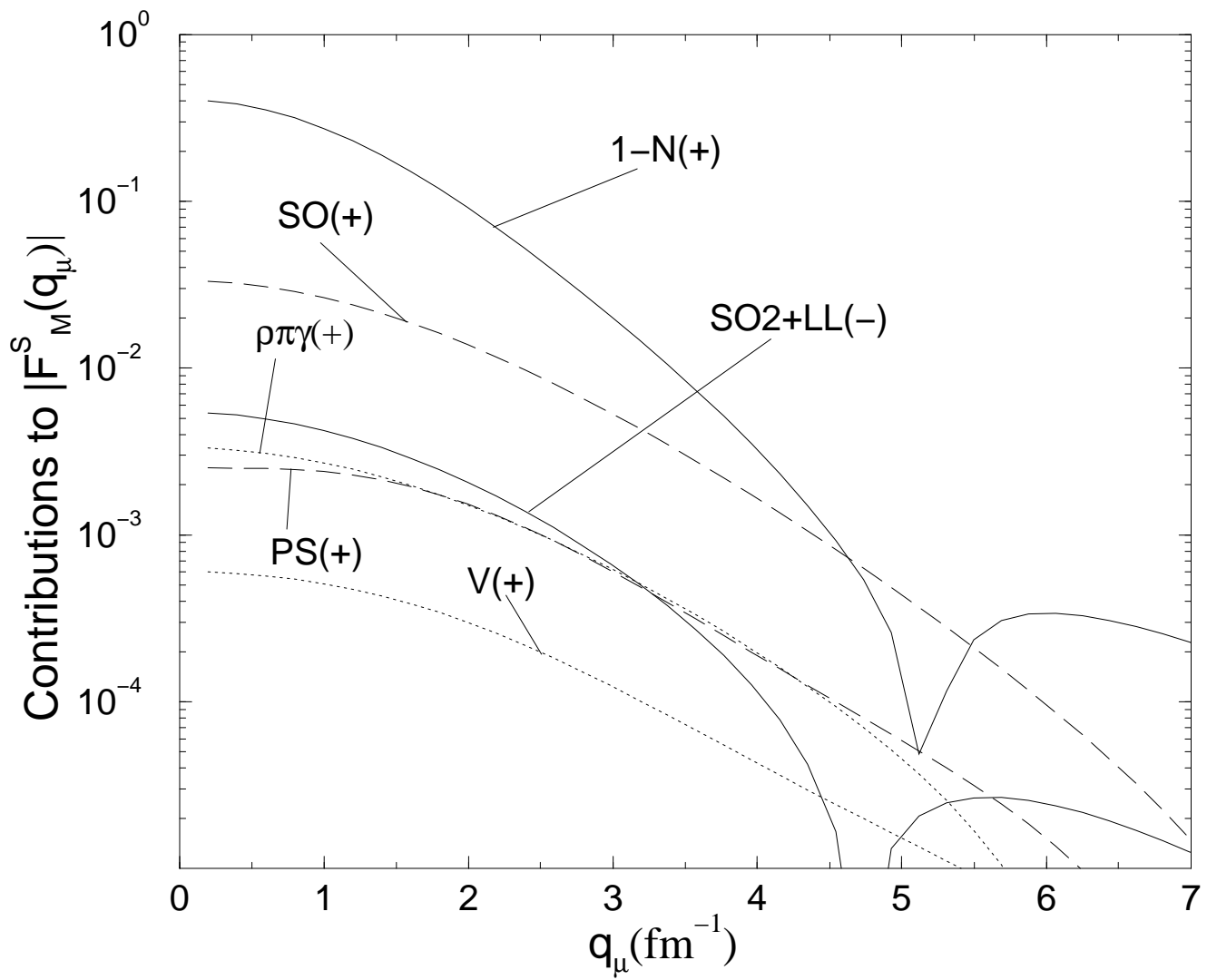
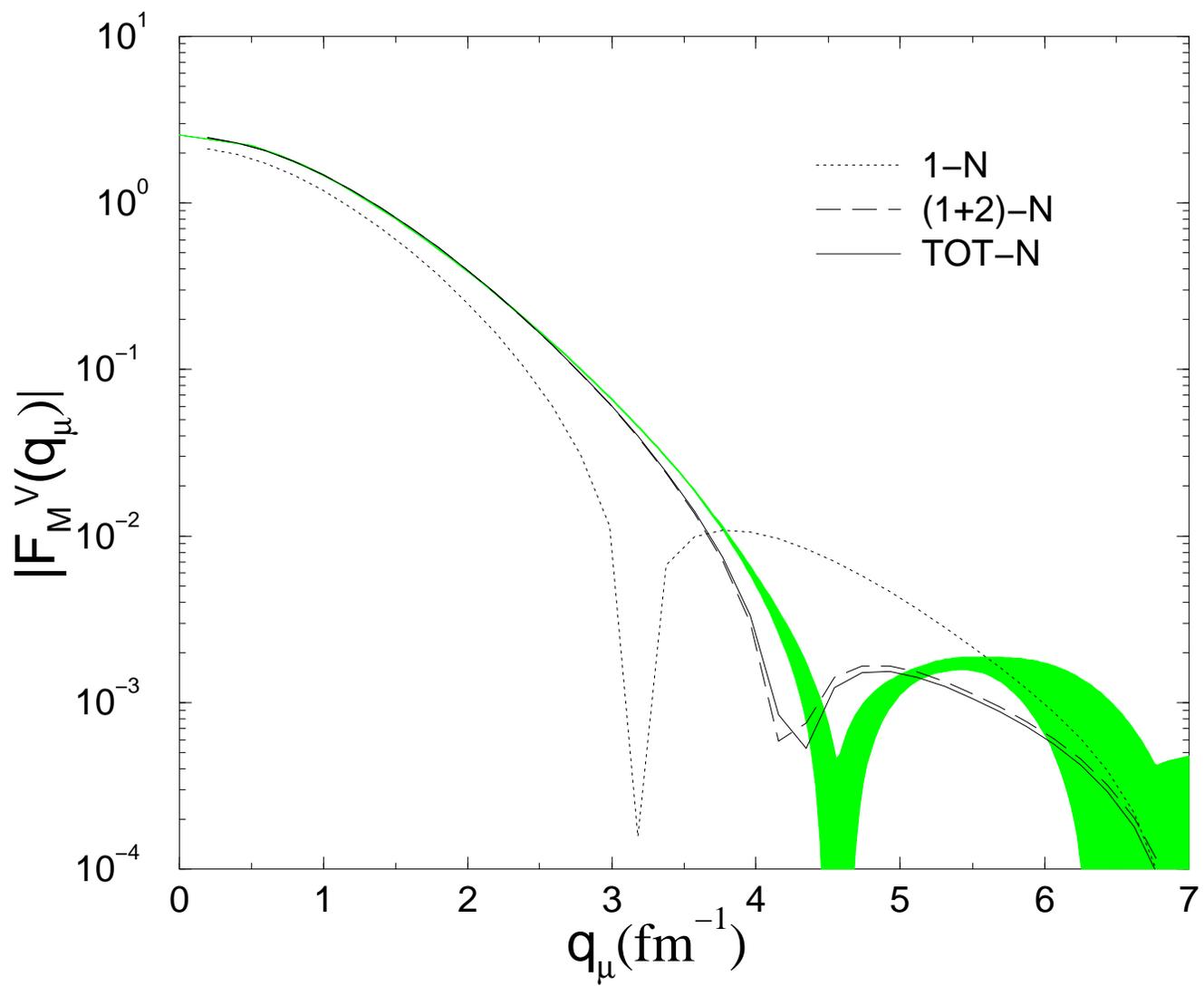
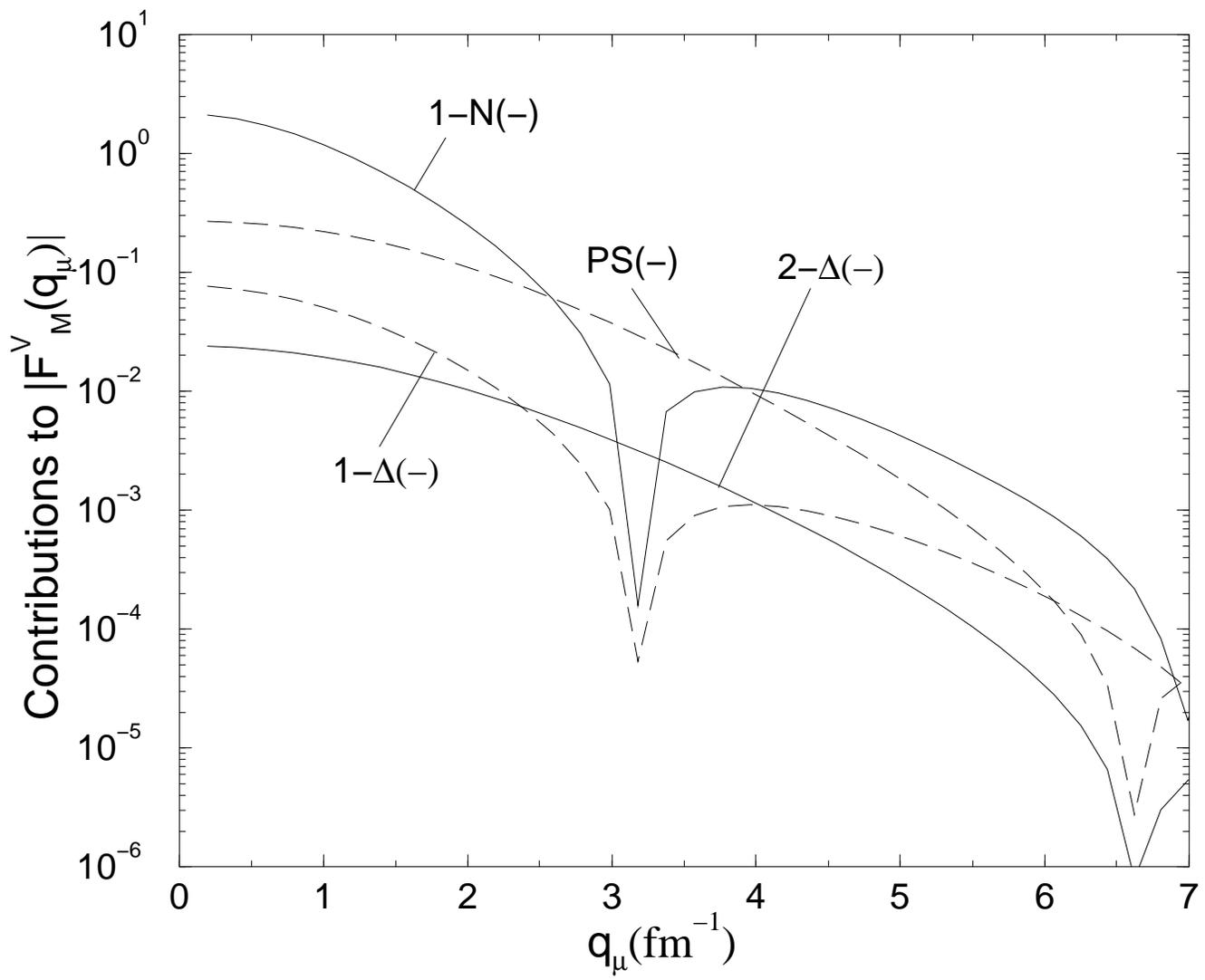
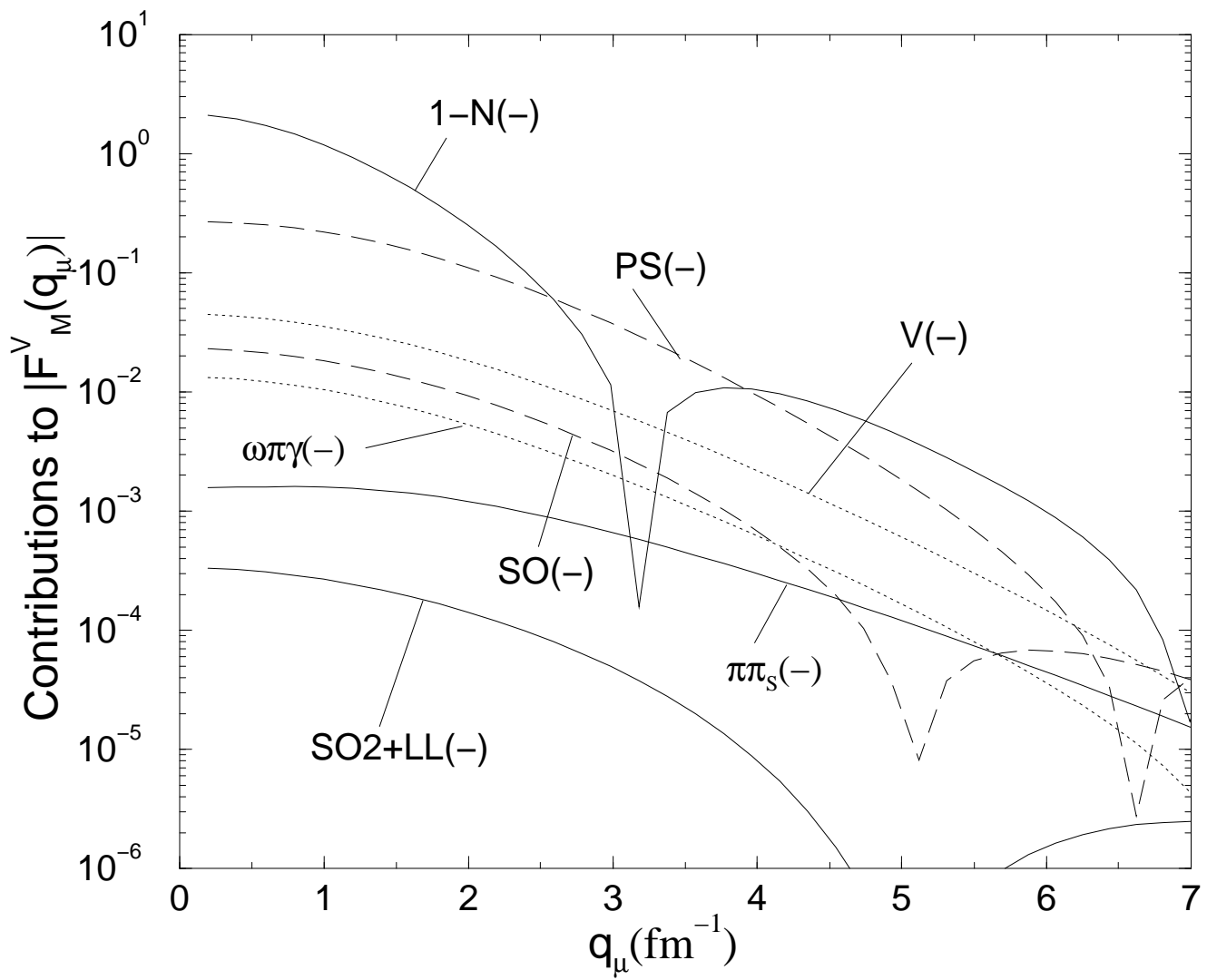
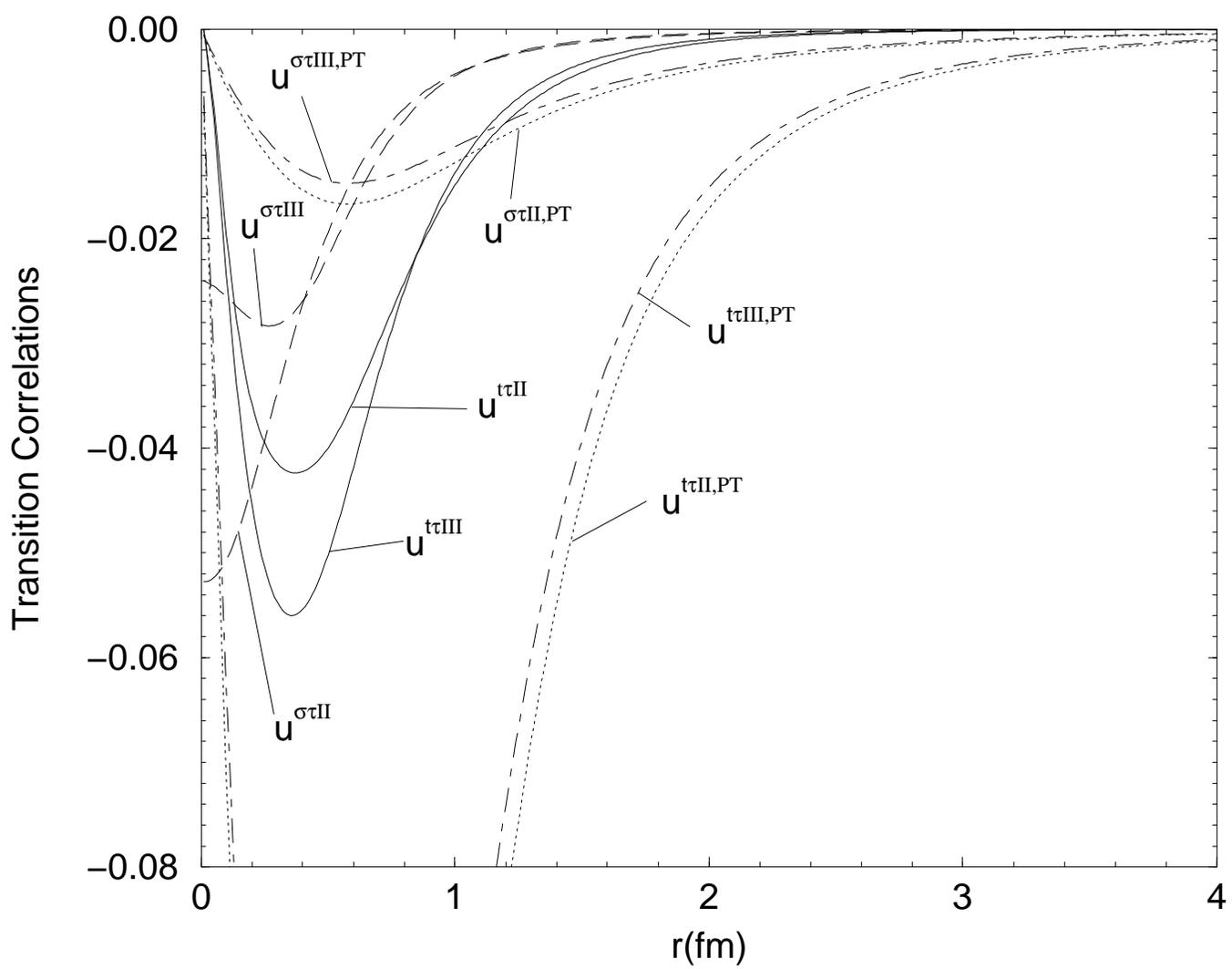
\end{document}